\documentclass[pra,nofootinbib,superscriptaddress,a4paper,twocolumn,10pt]{revtex4}

\usepackage{amsmath}
\usepackage{amsfonts}
\usepackage{amssymb}
\usepackage{amsthm}
\usepackage{mathrsfs}
\usepackage{dsfont}
\usepackage{esint}
\usepackage{graphicx}
\usepackage{physics}

\DeclareMathAlphabet{\mathpzc}{OT1}{pzc}{m}{it}

	\newcommand{\PropTo}{\propto}
	\newcommand{\AsymEq}{\sim}
	\newcommand{\ApproxEq}{\approx}

	\newcommand{\sKet}[1]{|#1\rangle}

	\newcommand{\sBK}[2]{\langle#1|#2\rangle}

	\newcommand{\V}[1]{\ensuremath{\boldsymbol{#1}}}			

	\newcommand{\mr}[1]{\mathrm{#1}}			

	\newcommand{\br}[1]{\left( #1 \right)}
	\newcommand{\brr}[1]{\left[ #1 \right]}
	
	\newcommand{\of}[1]{\!\br{#1}}
	\newcommand{\off}[1]{\!\brr{#1}}
	
	\newcommand{\sbr}[1]{( #1 )}
	\newcommand{\sbrr}[1]{[ #1 ]}
	
	\newcommand{\sof}[1]{\!\sbr{#1}}
	\newcommand{\soff}[1]{\!\sbrr{#1}}

	\newcommand{\Sum}[2]{\sum\limits_{#1}^{#2}}

	\newcommand{\Int}[3]{\int\limits_{#1}^{#2}\mr{d}#3\,}
	
	\newcommand{\sSum}[2]{\sum_{#1}^{#2}}

	\newcommand{\sInt}[3]{\int_{#1}^{#2}\mr{d}#3\,}

	\newcommand{\EA}[1]{\xpc{#1}}
	\newcommand{\Var}[1]{\mr{Var}\off{#1}}

	\newcommand{\xpc}[1]{\left\langle #1 \right\rangle}

	\newcommand{\sEA}[1]{\sxpc{#1}}
	\newcommand{\sVar}[1]{\mr{Var}\soff{#1}}
	
	\newcommand{\sxpc}[1]{\langle #1 \rangle}

	\newcommand{\Reals}{\ensuremath{\mathbb{R}} }

	\newcommand{\Landau}[1]{\mathpzc{O}\of{#1}}
	\newcommand{\sLandau}[1]{\mathpzc{O}\sof{#1}}
	\newcommand{\landau}[1]{\mathpzc{o}\of{#1}}

\newcommand{\PsiDet}{\psi_\mathrm{d}}
\newcommand{\PsiIn}{\psi_\mathrm{in}}

\newcommand{\Ham}{\hat{H}}

\newcommand{\FDP}{F}

\newcommand{\TDP}{P_\text{det}}

\newcommand{\Detect}{\hat{D}}

\newcommand{\RDet}{r_\text{d}}
\newcommand{\PSI}{\Psi}

\newcommand{\Bromwich}{\mathcal{B}}

\renewcommand{\TDP}{P_\text{d}}

\begin{document}
  \title{Quantization of the mean decay time for non-Hermitian quantum systems}
  \date{Manuscript of \today}
  \author{Felix Thiel}
  \email{thiel@posteo.de}
  \affiliation{Department of Physics, Institute of Nanotechnology and Advanced Materials, Bar-Ilan University, Ramat-Gan 52900, Israel}
  \affiliation{Physikalisches Institut, Albert-Ludwigs-Universit\"at Freiburg, Hermann-Herder-Str.~3, 79104 Freiburg, Germany}
  \author{David A. Kessler}
  \affiliation{Department of Physics, Institute of Nanotechnology and Advanced Materials, Bar-Ilan University, Ramat-Gan 52900, Israel}
  \author{Eli Barkai}
  \affiliation{Department of Physics, Institute of Nanotechnology and Advanced Materials, Bar-Ilan University, Ramat-Gan 52900, Israel}
  \begin{abstract}
    We show that the mean time, which a quantum particle needs to escape from a system to the environment, is quantized and independent from most dynamical details of the system.
    In particular, we consider a quantum system with a general Hermitian Hamiltonian $\Ham$ and one decay channel, through which probability dissipates to the environment with rate $\Gamma$.
    When the system is initially prepared exactly in the decay state, the mean decay time $\EA{T}$ is quantized and equal to $w/(2\Gamma)$.
    $w$ is the number of distinct energy levels, i.e. eigenvalues of $\Ham$, that have overlap with the decay state, and is also the winding number of a transform of the resolvent in the complex plane.
    Apart from the integer $w$, $\EA{T}$ is completely independent of the system's dynamics.
    The complete decay time distribution can be obtained from an electrostatic analogy and features rare events of very large dissipation times for parameter choices close to critical points, where $w$ changes, e.g. when a degeneracy is lifted.
    Experiments of insufficient observation time may thus measure a too small value of $w$.
    We discuss our findings in a disordered tight-binding model and in the two-level atom in a continuous-wave field.
  \end{abstract}

  \maketitle

  \section{Introduction}
    \begin{figure}
      \includegraphics[width=0.99\columnwidth]{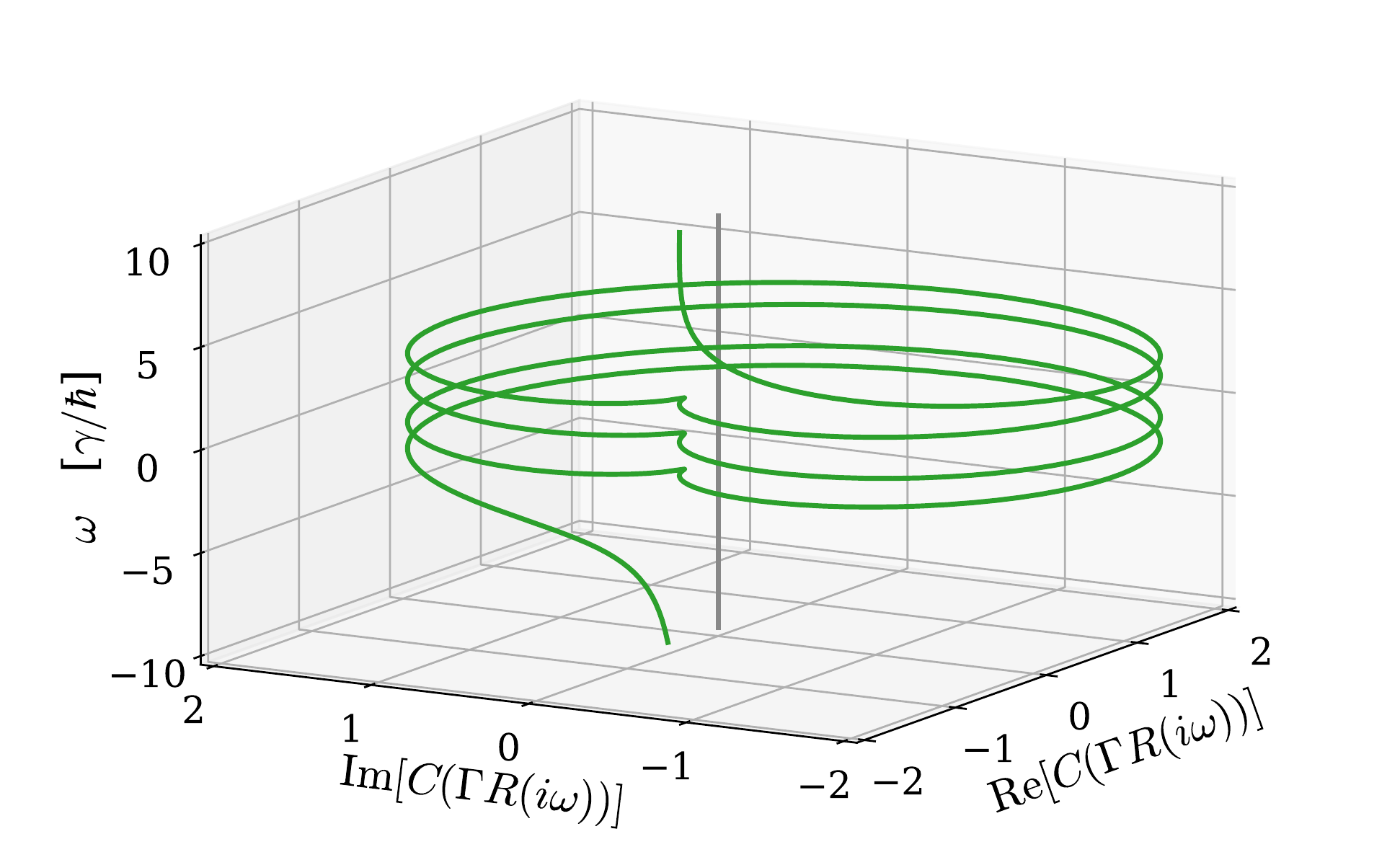}
      \caption{
        The map $C(\Gamma R(i\omega))$ maps the imaginary axis $\Bromwich = i \Reals$ to a curve that winds $-w$ times around the origin.
        This winding number determines $\EA{T}$ and equals the number of energy levels that have eigenstates non-orthogonal to $\ket{\PsiDet}$, as well as the number of stationary points in a related electrostatic potential, see Fig.~\ref{fig:Charge}.
        Here $w=4$ for the ring Hamiltonian, Eq.~\eqref{eq:TBHam} with $L=6$, $\epsilon=0$, and $\Gamma = 4.\gamma/\hbar$.
        $R(s)$ is the Hamiltonian's resolvent of Eq.~\eqref{eq:DefResolvent}.
        $C(z)$ is explained below Eq.~\eqref{eq:DefLogCurve}.
        \label{fig:Winding}
      }
    \end{figure}
    The quantization of certain observables is the very eponym of quantum theory.
    Topological and geometrical phases provide non-trivial mechanisms to this end \cite{Berry1984a, Zak1989a}.
    At the core of conventional quantum physics lies the Schr\"odinger equation with an Hermitian Hamiltonian $\Ham$.
    Yet, Hermiticity is not sacred and the ubiquitous exchange processes with the environment can be described with non-Hermitian terms \cite{Gamow1928a, Feshbach1958a, Ho1983a, Buchleitner1994a, Bender2007a, Moiseyev2011a}.
    The interplay between topology and non-Hermiticity is far from obvious.
    Non-Hermiticity may break topological quantization \cite{Philip2018a, Chen2018a}, but may as well lead to new quantized observables \cite{Rudner2009a}.
    A recent surge in the research of non-Hermitian systems, in particular of PT-symmetric systems \cite{Konotop2016a, El-Ganainy2018a} and of topological band theory \cite{Gong2018a, Ghatak2019a, Kawabata2019a}, together with their experimental realization \cite{Xu2017a, Rivet2018a, Xiao2019a, Lapp2019a, Li2019a} has brought much illumination upon the subject.

  \section{Model}
    This letter discusses a fundamental topic: the lifetime of a quantum state, or equivalently, the time until it dissipates, e.g. through emission of a photon.
    We here consider systems with a single dissipation channel $\ket{\PsiDet}$:
    \begin{equation}
      i \hbar \frac{\dd}{\dd t} \sKet{\psi\sof{t}}
      =
      \Ham \sKet{\psi\sof{t}} 
      - 
      i \hbar \Gamma
      \sKet{\PsiDet}\sBK{\PsiDet}{\psi\sof{t}}
      .
    \label{eq:Schroedi}
    \end{equation}
    The Hermitian Hamiltonian $\Ham$ mediates coherent transport and $\Gamma$ is the rate with which amplitude leaves the system.
    This system is relevant to quantum transport to state $\ket{\PsiDet}$ \cite{Caruso2009a, Agliari2010a, Muelken2011a, Krapivsky2014a, Novo2015a, Giusteri2015a}, to fluorescence with spontaneous emission from $\ket{\PsiDet}$ \cite{Plenio1998a}, and to the unraveling of master equations \cite{Gisin1992a, Meystre1988a, Dalibard1992a, Plenio1998a, Brun2002a}.

  \section{Result}
    We are interested in the random dissipation time $T$, which is the time of first spontaneous photon emission.
    The quantization of the mean recurrence time in stroboscopically measured systems, discovered by Gr\"unbaum, et al. \cite{Gruenbaum2013a}, and the connection \cite{Schulman1998a, Echanobe2008a, Facchi2008a, Muga2008a, Schaefer2014a, Dhar2015a, Dhar2015b, Elliott2016a, Mueller2017a, Lahiri2019a} between these and non-Hermitian systems in the limit of short intervals between measurements, suggest a search for a quantization in the latter as well.
    Indeed, we find that the mean dissipation time is quantized, provided $\ket{\psi(t=0)} = \ket{\PsiDet}$:
    \begin{equation}
      \EA{T}
      =
      \frac{w}{2\Gamma}
      .
    \label{eq:Quant}
    \end{equation}
    Besides $\Gamma$, $\EA{T}$ only depends on $w$, the number of distinct energy levels $E_l$ (i.e. the eigenvalues of $\Ham$), that have overlap with $\ket{\PsiDet}$, that means, which have eigenstates which are non-orthogonal to $\ket{\PsiDet}$.
    In other words, $w$ is the number of energy levels connected to the decay channel.
    $w$ can only be altered by a qualitative change of the energy spectrum.
    We show that it can be interpreted as a topological invariant, the winding number of a certain complex transformation of the resolvent $R(s) := \ev*{[s + i\Ham/\hbar]^{-1}}{\PsiDet}$.
    The quantization is a consequence of the discrete nature of the energy spectrum.
    Hence a ``change of topological class'' can be achieved by modifying either the decay channel, or the Hamiltonian $\Ham$, e.g. through lifting its degeneracies, as demonstrated later.

  \section{Quantization}
    The quantisation holds for a special initial condition, namely $\ket{\psi(t=0)} = \ket{\PsiDet}$, as mentioned. 
    In other cases, it is obvious from a physical point of view that the mean decay time depends on the separation between the initial and target state, and hence cannot be so elegant. 
    Therefore, to understand the behavior of $\EA{T}$, we need to solve Eq.~\eqref{eq:Schroedi} with the initial condition $\ket{\PsiDet}$.
    This can be achieved via a Laplace transform.
    Let the solution be $\ket{\psi(t)}$.
    Its squared norm is the system's survival probability that steadily decays to zero.
    The negative derivative of this norm $F(t) := - \dd /  \dd t \ip{\psi(t)} = 2 \Gamma\abs{\ip{\PsiDet}{\psi(t)}}^2$ describes the instantaneous decay rate or the probability density function of the decay time \footnote{
      The normalization of $F(t)$ is the total dissipation probability $\TDP = \sInt{0}{\infty}{t} F(t)$.
      Since the system is prepared in the decay state $\ket{\PsiDet}$ initially, the theory of invariant subspaces ensures $\TDP = 1$ \cite{Caruso2009a}.
      This is also confirmed from the small or large $\Gamma$ expansions of $F(t)$, see App.~\ref{app:TDP}.
    }.
    In App.~\ref{app:Formal}, we obtain the Laplace transform of ``the wave function''\footnote{
      The (non-Hermitian) wave function of the system is $\ket{\psi(t)}$, the solution Eq.~\eqref{eq:Schroedi}. 
      Since the specific component $\PSI(t) := \ip*{\PsiDet}{\psi(t)}$ is actually the only relevant part of the wave function for our purpose, we took the freedom of identifying it as ``the'' wave function.
    } $\PSI(t) := \ip{\PsiDet}{\psi(t)}$, namely $\PSI(s) := \sInt{0}{\infty}{t} e^{-st} \ip{\PsiDet}{\psi(t)}$.
    This is sufficient to obtain $\EA{T}$, which is the first moment of the distribution $F(t)$:
    \begin{equation}
      \EA{T}
      =
      \Int{0}{\infty}{t} t F(t) 
      =
      2\Gamma\int\limits_{\Bromwich}
      \frac{\dd s}{2\pi i}
      \underbrace{
        \PSI^*(-s) \qty(- \tfrac{\dd}{\dd s}) \PSI(s)
      }_{=: I(s)}
      .
    \label{eq:MeanComplexAnal}
    \end{equation}
    Here $f^*(s) = [f(s^*)]^*$, $z^*$ is the complex conjugate of $z$, and we have 
    used standard rules to write the time-domain integral as a Laplace-domain integral.
    The complex integral is taken along the Bromwich path $\Bromwich = \{ 0^+ + i\omega | \omega \in \Reals \}$ 
    which lies immediately to the right of the imaginary axis and gathers all residues of 
    poles in the complex left half plane.
    The integrand as denoted in Eq.~\eqref{eq:MeanComplexAnal} is called $I(s)$.
    Then,
    \begin{equation}
      \PSI(s)
      =
      \frac{R(s)}{1 + \Gamma R(s)}
      \qc
      R(s)
      :=
      \ev{\frac{1}{s + \frac{i}{\hbar}\Ham}}{\PsiDet}
      ,
    \label{eq:PSIofR}
    \end{equation}
    where $R(s)$ is the Hamiltonian's resolvent.
    One finds that $R^*(-s) = - R(s)$ for $s=i\omega$.
    This allows one to rewrite the integrand as a logarithmic derivative:
    \begin{equation}
      I(s)
      = 
      - \frac{1}{4\Gamma^2}
      \frac{\dd }{\dd s} \ln C\qty(\Gamma R(s))
      ,
    \label{eq:DefLogCurve}
    \end{equation}
    where $C(z) = M(z)e^{M(z)}$ and $M(z) = (z-1)/(z+1)$.
    The argument principle ensures that $2\Gamma \EA{T} = w$ is given by the winding number $-w$ of the path $C(\Gamma R(\Bromwich))$ around the origin, which is depicted in Fig.~\ref{fig:Winding}.

    To complete our argument, we need to specify the Hamiltonian.
    We assume rather generally that $\Ham = \sSum{l}{} E_l \sSum{m=1}{g_l} \dyad{E_{l,m}}$, and thus:
    \begin{equation}
      R(s)
      =
      \Sum{l=1}{w} \frac{p_l}{s + \frac{i}{\hbar} E_l}
      ,
    \label{eq:DefResolvent}
    \end{equation}
    where $E_l$ are the $g_l$-fold degenerate energy levels with eigenstates $\ket{E_{l,m}}$.
    $p_l := \sSum{m=1}{g_l} \abs{\ip*{E_{l,m}}{\PsiDet}}^2$ is the overlap of $\ket{\PsiDet}$ with the $E_l$-eigenspace, such that $\sSum{l=1}{w} p_l = 1$.
    Spectral components absent in $\ket{\PsiDet}$ are irrelevant.
    Eq.~\eqref{eq:DefResolvent} thus expresses our assumption that there are $w$ distinct energy levels that appear in $\ket{\PsiDet}$, and that the dissipation state is orthogonal to any part of the Hilbert space that belongs to a continuum of energy states.
    Any additional states that have access only to $\ket{\PsiDet}$, but are not considered ``part of the system or of the dynamics'' would lead to a modified effective dissipation strength $\Gamma$ and can be neglected.
    Later, we will also discuss the influence of energy states that are only weakly overlapping with the detection state.

    It turns out that $R(s)$ maps the imaginary axis $w$ times to itself.
    In fact each interval $[-iE_{l+1},-iE_l)$ as well as the outer segment $[-i\infty,-iE_w)\cup [-iE_0,i\infty)$ is mapped to the complete imaginary axis by $R(s)$ \footnote{Fig.~\ref{fig:Resolvent} in the appendix demonstrates this nicely.}.
    The M\"obius transform $M(z)$ maps $\Bromwich$ to the unit circle, which gets mapped to another curve circling the origin by $z e^z$.
    The direction is reversed, hence the winding number is $-w$, and $\EA{T}$ is related to this topological invariant of $C(\Gamma R(s))$.
    Thus Eq.~\eqref{eq:Quant} and $w$ is the number of distinct energy levels appearing in $\ket{\PsiDet}$.
    Application of the residue theorem to $I(s)$ gives the same result.

  \section{Two-Level system}
    \begin{figure}
      \includegraphics[width=0.99\columnwidth]{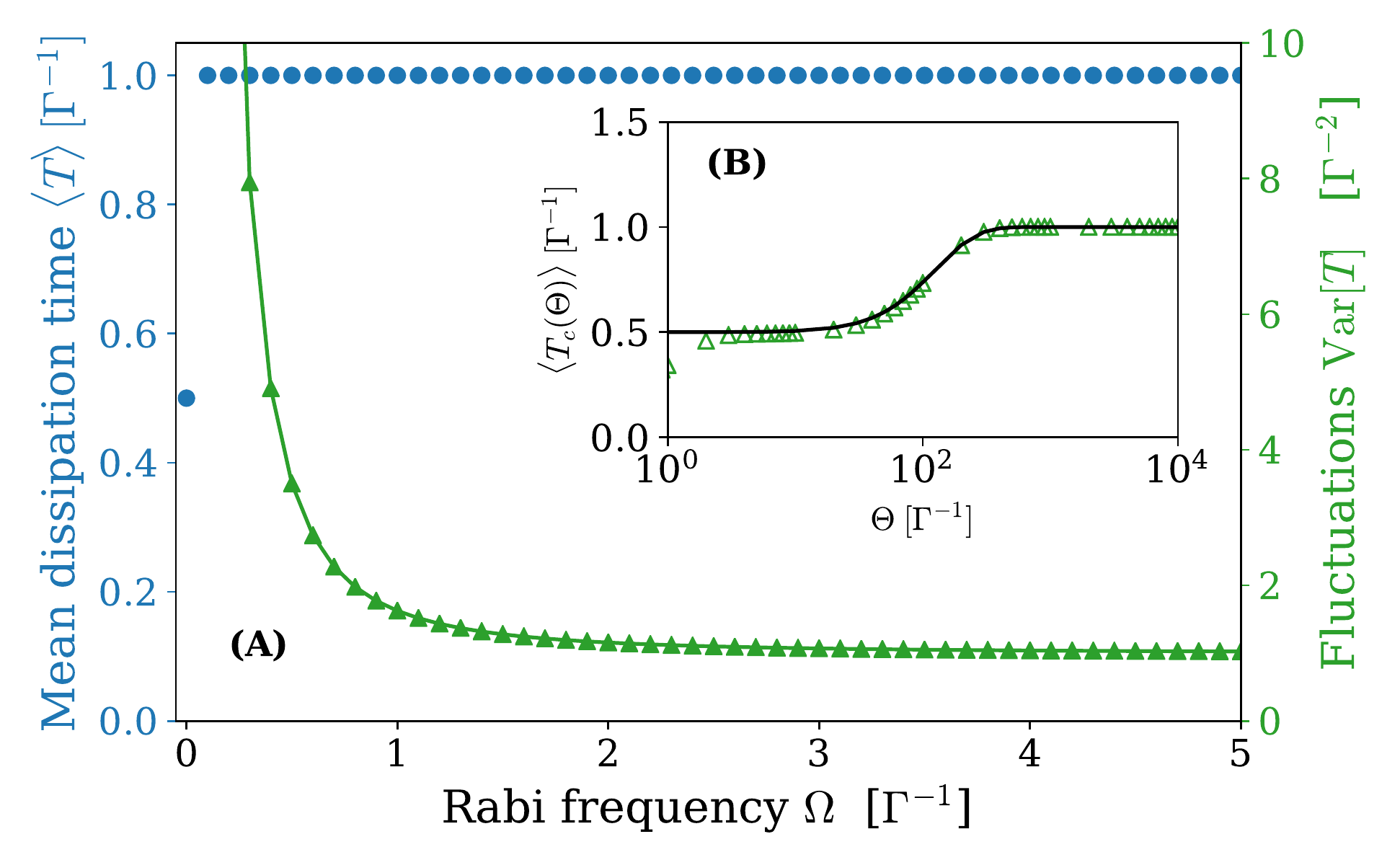}
      \caption{
        (A) Mean decay time (blue circles, left ordinate) and $\sVar{T}$ (green triangles, right ordinate) for the two-level system 
        with detuning $\delta = 0.5\Gamma$ ($w=2$).
        The mean decay time is quantized and stays constant except for vanishing Rabi frequency $\Omega$ (where $w=1$).
        The fluctuations, however, diverge in this limit.
        (B) Conditional mean dissipation time versus observation time $\Theta$ (see below) for the same parameters with $\Omega=0.1\Gamma^{-1}$.
        Full expression (symbols) and approximation of Eq.~\eqref{eq:TwoLevelCondMean} (solid line).
        Only for large $\Theta \gg \Omega^{-2}$, the small overlap can be resolved in $\sEA{T_c}$.
        \label{fig:TwoLevel}
      }
    \end{figure}
    Let us now consider an elementary quantum optics model, an atom with an excited ($e$) and a ground state ($g$) in a continuous-wave (cw) field close to resonance.
    Under the rotating wave approximation, the Schr\"odinger equation for this system reads \cite{Grynberg2010a}:
    \begin{equation}
      i
      \mqty( \dot{\psi}_e(t) \\ \dot{\psi}_g(t) )
      =
      \mqty( \frac{\delta}{2} & \Omega \\ \Omega & - \frac{\delta}{2} )
      \mqty( \psi_e(t) \\ \psi_g(t) )
      - i 
      \mqty( \Gamma & 0 \\ 0 & 0 )
      \mqty( \psi_e(t) \\ \psi_g(t) )
      .
    \label{eq:TwoLevel}
    \end{equation}
    Here, $\delta$ is the detuning, $\Omega$ is the Rabi frequency and $\Gamma$ is the inverse lifetime of the excited state.
    Our theory claims that the mean decay time depends only on $\Gamma$ and the number $w$ of energy levels whose eigenstates are non-orthogonal to $\ket{\PsiDet} = \ket{e} = (1, 0)^T$.
    Since the eigenstates of the Hermitian part of Eq.~\eqref{eq:TwoLevel} are given by $\ket{E_\pm} = N_\pm [ \ket{g} + ( \delta \pm \sqrt{\delta^2 + 4\Omega^2}) \ket{e}/2 ]$,
    with appropriate normalization $N_\pm$, we find $w=2$ and $\EA{T} = \Gamma^{-1}$ from Eq.~\eqref{eq:Quant}.
    The mean decay time is equal to the inverse life time.

    The exception occurs when the cw field is turned off and $\Omega$ vanishes.
    Then $p_- = \abs{\ip{\PsiDet}{E_-}}^2 = 0$ and $w$ drops to unity, discontinuously halving $\sEA{T}$.
    As this point is approached, for small but finite $\Omega$, the fluctuations $\sVar{T}$ of the decay time diverge, see Fig.~\ref{fig:TwoLevel}(A), see App.~\ref{app:TwoLevel}.
    This divergence is due to the tiny portion of the wave function that escapes to the ground state before it decays.
    There it spends a long time before returning to $\ket{\PsiDet} = \ket{e}$, where it can dissipate.

    So far, we focused on the mean of $T$ and only briefly on its fluctuations in an example.
    A complete picture from the investigation of $T$'s whole distribution shows that large outliers and divergent fluctuations are a signature of every critical point where the topological number changes.

  \section{Dissipation time distribution}
    $\PSI(s)$ has $w$ simple poles $s_{p,l}$ in the left-half plane that give the wave function via the inverse Laplace transform and the residue theorem:
    \begin{equation}
      \PSI(t)
      :=
      \Sum{l=1}{w}
      \underset{s\to s_{p,l}}{\mathrm{Res}}\; \PSI(s) e^{st}
      =
      \frac{1}{\Gamma}
      \Sum{l=1}{w}
      r_l
      e^{s_{p,l}t}
      ,
    \label{eq:TimeDomainPSI}
    \end{equation}
    where $r_l = \Gamma \mathrm{Res}_{s_{p,l}} \PSI(s) = R(s_{p,l})/R'(s_{p,l})$ is the ``residual coefficient''.
    This sum identifies $F(t) = 2\Gamma \abs{\Psi(t)}^2$ as a sum of exponential modes with decay rates $-\Re[s_{p,l} + s_{p,l'}]$, oscillating with frequencies $\Im[s_{p,l} - s_{p,l'}]$.
    Integrating the resulting $F(t)$ against $t^m$ yields the moment $\EA{T^m}$ in terms of the poles and residuals:
    \begin{equation}
      \EA{T^m}
      =
      \frac{2}{\Gamma} \sSum{l,l' = 1}{w} \frac{m! r_l r_{l'}^*}{(-s_{p,l} - s_{p,l'}^*)^{m+1}}
      .
    \label{eq:HighMoments}
    \end{equation}
    Clearly, poles close to the imaginary axis give slowly decaying modes in $\PSI(t)$ and also dominate the sum in Eq.~\eqref{eq:HighMoments}.
    Knowing the poles $s_{p,l}$ of $\PSI(s)$ thus determines the full distribution and all moments of the dissipation time.
    They also appear as second order poles in the integrand $I(s)$ above.
    Finding the poles is generally a hard task, which is greatly facilitated by the following analogy, particularly close to the critical points, when $w$ changes.

  \section{Electrostatic analogy}
    \begin{figure}
    \centering
      \includegraphics[width=0.99\columnwidth]{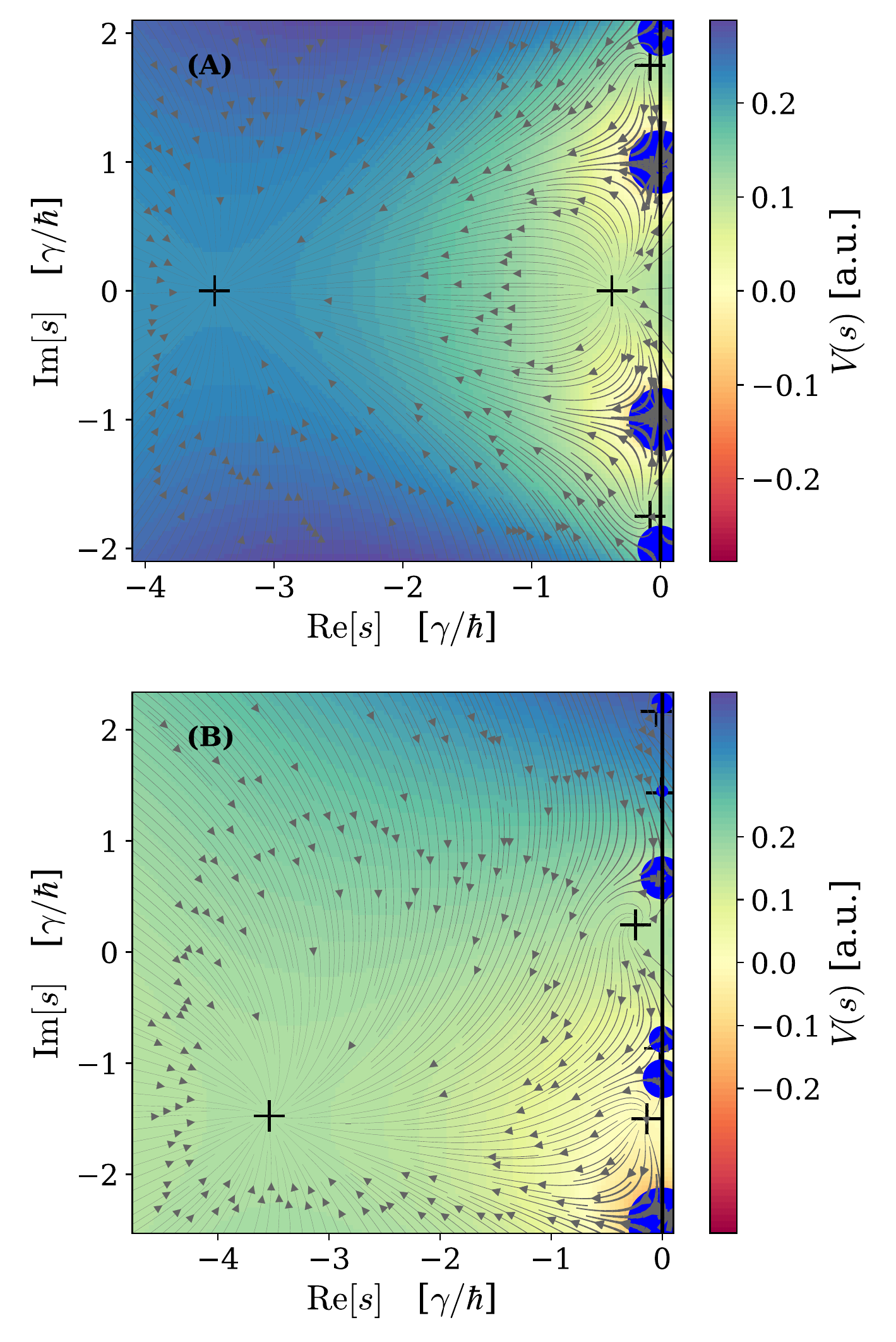}
      \caption{
        The poles $s_{p,l}$ ($+$) are the stationary points of a 2D-electrostatic field $V(s)$, [Eq.~\eqref{eq:Coulomb}, heat map, stream lines equal gradient], which consists of a constant force and point charges at $-E_l/\hbar$ on the imaginary axis of magnitude $p_l$ (blue circles).
        Since all charges are positive, we have $w$ stationary points from electrostatics.
        The pole on the far left corresponds to a fast, quickly decaying mode.
        Stationary points can be found close to weak charges and pairs of close charges, see Appendix~\ref{app:Charge}.
        Both figures are for the Hamiltonian \eqref{eq:TBHam} with $L=6$, $\Gamma = 4\gamma/\hbar$ and $\epsilon = 0$ (A) and $\epsilon=1\gamma$ (B), respectively.
        \label{fig:Charge}
      }
    \end{figure}
    Consider the following 2D-electrostatic potential:
    \begin{equation}
      V(x,y) 
      := 
      \frac{x}{\Gamma}
      +
      \Sum{l=1}{w} p_l \ln \frac{1}{\Gamma}\sqrt{x^2 + (y+\tfrac{E_l}{\hbar})^2}
      ,
    \label{eq:Coulomb}
    \end{equation}
    which is constructed from a $\Gamma$-dependent constant force and $w$ 2D-point charges of magnitude $p_l$ at the positions $(0,-E_l/\hbar)$.
    The potential has $w$ stationary points defined by $\grad_{(x,y)} V = \V{0}$.
    Writing $s = x+iy$ reveals that $R(s) + 1/\Gamma = \pdv{V}{x} + \pdv{V}{iy} = (1, -i) \cdot \grad V$, such that the poles $s_{p,l}$ of $\PSI(s)$, defined by Eq.~\eqref{eq:PSIofR}, must be equal to the potential's stationary points (similarly to \cite{Gruenbaum2013a}).
    This picture enables one to draw many general conclusions about the position of the poles.
    For instance they all must have negative real part due to the constant force.
    Thus $F(t)$ is bounded.
    The positions of the stationary points can be predicted from an electrostatic force balance, see App.~\ref{app:Charge} and also Refs.~\cite{Yin2019a, Liu2020a}.
    In several limits \footnote{In App.~\ref{app:Charge} we consider the limits of vanishing charges, of approximate degeneracy, as well as vanishing and diverging dissipation rate.}, some stationary poles move close to the imaginary axis, see Fig.~\ref{fig:Charge}, and dominate the dissipation time statistics.

    This helps understanding the divergent fluctuations in the two-level system:
    One charge $p_- \AsymEq (\Omega/\delta)^2$ vanishes with diminishing Rabi frequency.
    This gives a small decay rate $\Re[s_{p,-}] \PropTo -p_-$ coefficient $r_- \PropTo p_-$ and eventually a large $\sVar{T} \PropTo 1/p_- \PropTo \Omega^{-2}$ from Eq.~\eqref{eq:HighMoments}, see Fig.~\ref{fig:TwoLevel} and App.~\ref{app:TwoLevel}.
    Next, we discuss changing $w$ by lifting degeneracies.

  \section{Tight-binding model}
    \begin{figure}
      \includegraphics[width=0.99\columnwidth]{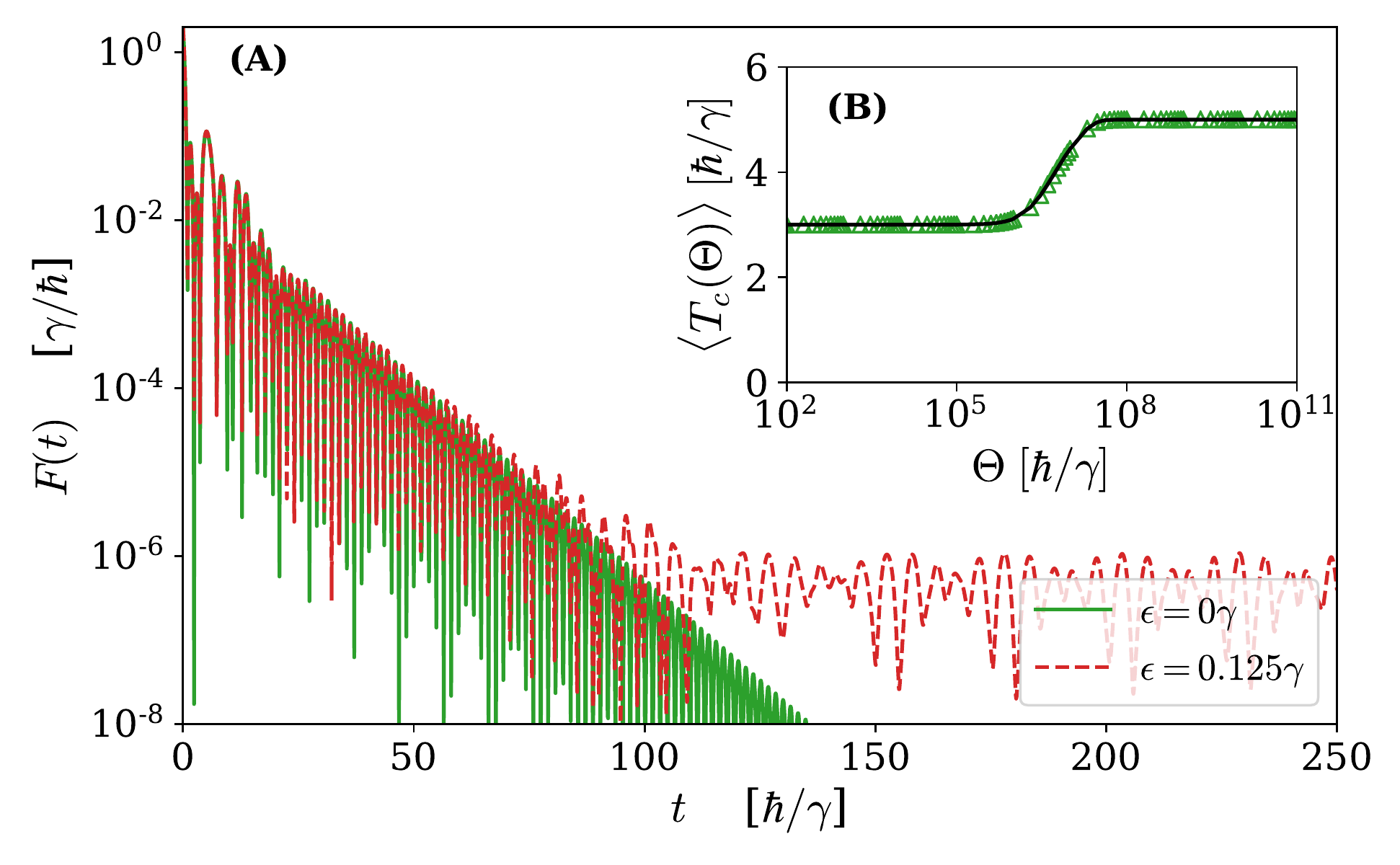}
      \caption{
        (A): Decay time distribution $F(t)$ for the disordered tight-binding model with $\Gamma = \gamma/\hbar$, $L=10$ and $\epsilon =0$ (solid green line), or $\epsilon=0.125\gamma$ (dashed red line).
        Introduction of disorder only changes the slow tails of $F(t)$ for large times.
        Before some crossover time both distributions are practically indistinguishable.
        (B): The conditional mean dissipation time $\EA{T_c(\Theta)}$ against observation time $\Theta$ for the same model with $\epsilon=1/1024\gamma$.
        For short observation times degenerate energy levels can not be resolved and appear as one, resulting in a coarse-grained quantization.
        Only for large $\Theta\gg\epsilon^2$, the true value $\EA{T_c} = \EA{T} = w/(2\Gamma)$ emerges, see Eq.~\eqref{eq:CondMeanPerturb}.
        \label{fig:RingDis}
      }
    \end{figure}
    The tight-binding Hamiltonian is a basic transport model with a degenerate spectrum:
    \begin{equation}
      \Ham
      =
      \Sum{x=1}{L} \qty( \epsilon_x \dyad{x} - \gamma \dyad{x}{x+1} - \gamma \dyad{x+1}{x} )
      .
    \label{eq:TBHam}
    \end{equation}
    Here $\gamma$ is the hopping energy, and $\ket{x}$ are position eigenstates on a ring with $L$ sites and periodic boundary conditions $\ket{x+L} = \ket{x}$, where $L$ is even.
    The random on-site energies $\epsilon_x$ are i.i.d. variables drawn uniformly from $[-\epsilon, \epsilon]$.
    We equip one site ($\RDet$) with a sink and set $\ket{\PsiDet} = \ket{\RDet}$ in Eq.~\eqref{eq:Schroedi}.
    For vanishing $\epsilon$, $\Ham$ has $(L-2)/2$ pairs of degenerate energy levels, that split up for $\epsilon>0$.
    Clearly, there is a transition from $w = (L+2)/2$ to $w=L$ at $\epsilon=0$.
    However, the drastic change in the mean is accompanied with seemingly moderate adjustments in the distribution $F(t)$ when disorder is turned on.
    After some large crossover time, disorder manifests in very slowly decaying tails, see Fig.~\ref{fig:RingDis}(A).
    A similar crossover effect was reported in connection to a magnetic field \cite{Stupid2019a}.

    When $\epsilon>0$ is non-zero, the degeneracy of the ring-Hamiltonian's energy levels is lifted, such that several charges split up into pairs of distance $\sLandau{\epsilon}$.
    Close to each pair on the imaginary axis, a new stationary point emerges with $\Re[s_{p,l}] \PropTo - \epsilon^2$.
    The corresponding residual coefficient is also small $r_l \PropTo \epsilon^2$, see App.~\ref{app:Charge}, so that slow modes are only present in the tails of $F(t)$.
    Due to these heavy tails, the variance diverges like $\epsilon^{-2}$, see Eq.~\eqref{eq:HighMoments}.
    This is a signature of the critical point, where $w$ changes, but makes the measurement of $\EA{T}$ difficult.

  \section{Finite observation times}
    The culprit are rare outliers with very large values of $T$, that dominate the statistics and are problematic for realistic experiments of finite duration.
    A realistic measurement of $T$ has a cut-off time $\Theta$, and later dissipation events are not recorded.
    The {\em conditional} dissipation time $T_c(\Theta)$ takes only events smaller than $\Theta$ into account.
    Its mean is given by:
    \begin{equation}
      \EA{T_c(\Theta)} 
      :=
      \sInt{0}{\Theta}{t} t F(t)
      \bigg/
      \sInt{0}{\Theta}{t} F(t)
      .
    \label{eq:}
    \end{equation}
    This dramatically distorts the mean dissipation time close to critical points, where $w$ changes.
    $\sEA{T_c(\Theta)}$ can be expressed in terms of the poles $s_{p,l}$ and coefficients $r_l$, similar to Eq.~\eqref{eq:HighMoments}, see App.~\ref{app:FinObs}.
    For the two-level system with small $\Omega$ one finds:
    \begin{equation}
      \sEA{T_c(\Theta)}
      \AsymEq
      \frac{1}{2\Gamma} \qty[
        2 - 
        \qty( 1 + \frac{2\Gamma \Omega^2\Theta}{\Gamma^2 + \delta^2} ) 
        e^{-\frac{2\Gamma \Omega^2\Theta}{\Gamma^2 + \delta^2}}
      ]
      ,
    \label{eq:TwoLevelCondMean}
    \end{equation}
    for large $\Theta$ and small $\Omega$.
    The result is plotted in Fig.~\ref{fig:TwoLevel}(B).
    It exhibits a transition between the true ($w=2$) and the apparent ($w_\text{app}=1$) result due to the small charge $p_- \PropTo \Omega^2$.
    The crossover is controlled by the product $\Omega^2\Theta$, that can assume small, intermediary, or large values.
    The same is observed in the ring model, where pairs of energy levels split up when $\epsilon>0$, leading to a crossover from $w_\text{app} = 1 + L/2$ to $w=L$, see Fig.~\ref{fig:RingDis}(B).
    In general, when a system features $w-w_\text{app}$ poles that are $\sLandau{\varepsilon}$-close to the imaginary axis,\footnote{
      This is the case when $w-w_\text{app}$ charges are small of order $\sLandau{\varepsilon}$ or when a $w-w_\text{app}+1$-fold degenerate energy level splits under a perturbation of strength $\sLandau{\sqrt{\varepsilon}}$.
    }, then, as we show in App.~\ref{app:FinObs}, $\sEA{T_c(\Theta)}$ will behave like:
    \begin{equation}
      \EA{T_c(\Theta)}
      \AsymEq
      \frac{w_\text{app}}{2\Gamma} + \frac{w-w_\text{app}}{2\Gamma} g(\varepsilon\Theta)
      \qc \varepsilon \to0
      \qc \Theta\to \infty
      ,
    \label{eq:CondMeanPerturb}
    \end{equation}
    where $g(x)$ is a monotonic scaling function with $g(0) = 0$ and $g(\infty) = 1$, computed from $s_{p,l}$ and $r_l$, that mediates the transition between $w_\text{app}$ and $w$.
    Therefore, $\sEA{T_c(\Theta)}$ can only resolve energy differences of order $1/\sqrt{\Theta}$ and small overlaps of order $1/\Theta$.
    For smaller measurement times the energy levels will appear degenerate or missing, respectively, see Figs.~\ref{fig:TwoLevel}(B) and \ref{fig:RingDis}(B).
    Exactly the same behavior occurs in other critical scenarios, where $w$ changes, see App.~\ref{app:FinObs}.

  \section{Discussion and Summary}
    Our most remarkable result is how $\EA{T}$ completely lacks any dependence on the system's dynamical details.
    This relies critically on the special preparation state $\ket{\psi(t=0)} = \ket{\PsiDet}$ and the finite dimensionality.
    Violating these conditions break quantization to variable degrees as we discuss shortly in App.~\ref{app:Arr} and in more detail in a longer publication \cite{Thiel2020b}.

    As mentioned in the Introduction, the quantization we encounter here is deeply connected to the quantization of $\EA{T}$ in systems subject to periodic strong measurements. 
    Such systems in the Zeno limit of small period can be mapped to a strongly dissipative ($\Gamma\to\infty$) non-Hermitian system \cite{Schulman1998a, Echanobe2008a, Facchi2008a, Muga2008a, Schaefer2014a, Dhar2015a, Dhar2015b, Elliott2016a, Mueller2017a, Lahiri2019a}.
    The topological nature of the quantization is what guarantees its general validity for all values of damping.
    In that sense, the topological effects are robust.
    They are found far and close to the Zeno limit, for both, projective measurements and for non-Hermitian modeling. 

    Our assumption on a {\em single} dissipation channel is not crucial.
    The above mentioned quantization in repeatedly measured systems survives adding more channels \cite{Bourgain2014a}.
    App.~\ref{app:MultiDiss} provides preliminary numerical evidence that quantization also holds in the multi-channel case for general non-Hermitian systems.
    Further work in this direction is clearly warranted.

    Time-resolved fluorescence spectroscopy can be employed to measure the dissipation time.
    A suitable transition between an atomic/molecular ground state and some excited state is isolated.
    A short $\pi$-pulse prepares the system in the excited state, from which it spontaneously decays under fluorescent emission.
    $T$ is the time between the $\pi$-pulse and the detection of the fluorescence photon and is measured using time-correlated single photon counting.
    $T$ should be compared present or absent of additional Rabi driving.
    Alternatively, cold-atom experiments can verify our findings.
    In Ref.~\cite{Li2019a} the two hyper-fine levels of a $\mathrm{{}^6Li}$ gas are coupled via a radio-frequency field.
    A resonant optical beam is used to move atoms from one of the levels to a third excited level, thus simulating dissipation.
    Both examples implement the aforementioned two-level system.
    More complicated systems can be engineered with ``synthetic lattices'' \cite{Lapp2019a}.
    Measurements of the population in the principal lattice yield $F(t)$.

    We discussed the topological effect in the mean dissipation time in two finite-dimensional systems.
    Requiring a realistic system to be finite-dimensional (e.g. two-level) may appear questionable at first sight sight.
    Other atomic states may become relevant and affect the dissipation time statistics, which have been neglected in the model; for example almost degenerate energy levels or states with a very small overlap.
    However, as we have demonstrated their influence manifests only in rare outliers in the regime where $T$ is tremendously large.
    They are invisible in experiments with a large, but not infinite, observation time.
    The existence of these states constitutes a upper limit to the temporal validity of our theory.
    Thus for all practical reasons, the finite-level approximation is excellent.
    
    We have investigated the dissipation time $T$  through a single channel $\ket{\PsiDet}$ in an otherwise Hermitian system, demonstrating how its mean is quantized and how its fluctuations explode close to critical points where the topological number $w$ changes.
    We have discussed our findings in two examples and how to implement our model experimentally in cold-atoms or fluorescence experiments.

  \begin{acknowledgments}
    Felix Thiel thanks DFG (Germany) to support him under grants TH 2192/1-1 and TH 2192/2-1.
    He thanks Jonathan Ruhman for insightful discussions and Andreas Buchleitner for his hospitality at University of Freiburg.
    The support of Israel Science Foundation's grant 1898/17 is acknowledged.
  \end{acknowledgments}

  \appendix
  \numberwithin{equation}{section}
  \section{Derivation of the formal solution}
  \label{app:Formal}
    In this section, we derive the formal solution of Eq.~(1).

    This derivation closely follows Ref.~\cite{Krapivsky2014a}.
    Apply a Laplace transform to both sides of Eq.~(1).
    We write $\ket{\psi(s)} = \mathcal{L}\qty{\ket{\psi(t)};s} := \sInt{0}{\infty}{t} e^{-st}\ket{\psi(t)}$ 
    and use the initial condition $\ket{\psi(t=0)} = \ket{\PsiDet}$:
    \begin{equation}
      i \hbar [ s \ket{\psi(s)} - \ket{\PsiDet} ]
      =
      \Ham \ket{\psi(s)}
      -
      i \hbar \Gamma \ket{\PsiDet}
      \ip{\PsiDet}{\psi(s)}
      .
    \label{eq:}
    \end{equation}
    Rearranging the equation yields:
    \begin{equation}
      \ket{\psi(s)}
      =
      \qty[ s + \frac{i}{\hbar} \Ham ]^{-1}
      \qty[
        \ket{\PsiDet}
        -
        \Gamma \ket{\PsiDet}
        \ip{\PsiDet}{\psi(s)}
      ]
      .
    \label{eq:}
    \end{equation}
    When the equation is multiplied with $\bra{\PsiDet}$ from the left and solved
    for $\ip{\PsiDet}{\psi(s)} = \PSI(s)$, Eq.~(4) is obtained.

    To see how $\FDP(t)$ and $\PSI(t)$ are related, one considers the survival 
    probability, whose negative derivative is equal to $\FDP(t)$.
    The survival probability is equal to the not yet decayed probability, i.e. 
    equal to the norm of the state $\ket{\psi(t)}$.
    We find:
    \begin{align}
      \FDP(t) 
      = &
      - \dv{t} \ip{\psi(t)}{\psi(t)}
      \\ = & \nonumber
      - \qty(\dv{t} \bra{\psi(t)}) \ket{\psi(t)}
      - \bra{\psi(t)} \dv{t} \ket{\psi(t)}
      \\ = & \nonumber
      - \ev*{\qty[
        \frac{i}{\hbar}\Ham 
        - \frac{i}{\hbar}\Ham 
        - 2\Gamma \dyad{\PsiDet}
      ]}{\psi(t)}
      .
    \label{eq:}
    \end{align}
    Hence $F(t) = 2\Gamma \abs*{\PSI(t)}^2$ as stated in the main text.

    To obtain $\EA{T}$ in Laplace domain, we write 
    \begin{equation}
      \EA{T}
      =
      \lim_{\sigma\to0}
      2 \Gamma
      \Int{0}{\infty}{t} e^{-\sigma t} [\PSI(t)]^* [t \PSI(t)]
    \label{eq:}
    \end{equation}
    in terms of a Laplace transform.
    We have $[\PSI(t)]^* \to \PSI^*(s)$ and $[t \PSI(t)] \to (- \dd / \dd s) \PSI(s)$.
    Next we use the fact that products in time domain transform to convolutions in image domain:
    \begin{equation}
      \EA{T}
      =
      \lim_{\sigma\to0}
      2\Gamma 
      \int\limits_{\Bromwich}
      \frac{\dd s}{2\pi i }
      \PSI^*(\sigma - s) \qty( - \frac{\dd }{\dd s})\PSI(s)
      ,
    \label{eq:}
    \end{equation}
    where the integral only considers the poles of $\PSI(s)$.
    Taking the limit yields Eq.~(3).

  \section{The two-level system}
  \label{app:TwoLevel}
    The Schr\"odinger equation is given by Eq.~(8) of the main text.
    The energy levels (i.e. the eigenvalues of $\Ham$) are:
    \begin{equation}
      E_\pm 
      =
      \pm \frac{\hbar\delta}{2} \sqrt{ 1 + x^2 }
    \label{eq:}
    \end{equation}
    The corresponding eigenstates are:
    \begin{equation}
      \ket{E_\pm}
      =
      \frac{1}{\sqrt{2\qty( 1 + x^2 \pm x\sqrt{1+x^2} )}}
      \mqty( 1 \pm \sqrt{1 + x^2 }  \\ 1)
      ,
    \label{eq:}
    \end{equation}
    where $x := 2\Omega/\delta$.
    From here we obtain with $\ket{\PsiDet} = \ket{e} = (1 , 0 )^T$ the overlaps $p_\pm = \abs{\ip{e}{E_\pm}}^2$.
    These in turn give the resolvent:
    \begin{equation}
      R(s)
      =
      \frac{\abs{\ip*{\PsiDet}{E_+}}^2}{s + \frac{i}{\hbar}E_+}
      +
      \frac{\abs{\ip*{\PsiDet}{E_-}}^2}{s + \frac{i}{\hbar}E_-}
      =
      \frac{s - i\frac{\delta}{2}}{s^2 + \frac{\delta^2}{4} + \Omega^2}
      ,
    \label{eq:}
    \end{equation}
    which yields two simple poles $R(s_{p,\pm}) = -1/\Gamma$:
    \begin{equation}
      s_{p,\pm}
      =
      -
      \qty{
        \frac{\Gamma}{2}
        \pm
        \frac{\Gamma + i \delta}{2}
        \sqrt{ 1 - y^2 }
      }
      ,
    \label{eq:TwoLevelPoles}
    \end{equation}
    where $y = 2\Omega/(\Gamma + i\delta)$.
    The residual coefficients are:
    \begin{equation}
      r_\pm
      =
      \frac{R(s_{p,\pm})}{R'(s_{p,\pm})}
      =
      - \frac{\Gamma}{2} \frac{y^2}{1 - y^2 \mp \sqrt{1-y^2}}
      .
    \label{eq:TwoLevelResCoeffs}
    \end{equation}

    Using the poles and the residual coefficients, we can write down the wave function $\PSI(t)$ from Eq.~\eqref{eq:TimeDomainPSI} of the main text.
    The normalization $\TDP = \sInt{0}{\infty}{t} F(t)$ of the dissipation time distribution $F(t) = 2\Gamma \abs*{\PSI(t)}^2$ and the moments are obtained from Eq.~\eqref{eq:HighMoments} of the main text.
    We find:
    \begin{align}
      \TDP = 1 
      \qc
      \EA{T}
      =
      \frac{1}{\Gamma}
      \qc
      \Var{T}
      =
      \frac{\Gamma^2 + \delta^2 + 2\Omega^2}{2 \Gamma^2 \Omega^2}
      ,
    \label{eq:MomentsTwoLevel}
    \end{align}
    where of course $\sVar{T} = \sEA{T^2} - \sEA{T}^2$.
    For small Rabi frequency, we find: $\sVar{T} \AsymEq (\Gamma^2 + \delta^2)/(2\Gamma^2 \Omega^2)$.
    These are the quantities plotted in Fig.~2.

    An exception occurs for vanishing Rabi frequency.
    Then the decay state is an eigenstate of $\Ham$ and $w$ is equal to one.
    The resolvent reads $R(s) = 1/(s + i \delta/2)$ and there is only one pole $s_p = - \Gamma - i \delta/2$.
    In this case, we find
    \begin{align}
      \TDP = 1
      \qc
      \EA{T}
      =
      \frac{1}{2\Gamma}
      \qc
      \Var{T}
      = 
      \frac{1}{4\Gamma^2}
      .
    \end{align}

  \section{Charge theory}
  \label{app:Charge}
    Here, we investigate the position of the poles $s_{p,l}$ in several limits using the electrostatic analogy.
    This treatment parallels the one of Refs.~\cite{Yin2019a, Liu2020a}.
    
    The poles are the stationary points of the electrostatic potential of Eq.~\eqref{eq:Coulomb}.
    The corresponding electrostatic force is denoted as $E_\text{tot}(s;\varepsilon)$, where $\varepsilon$ denotes some small parameter that is varied.
    Let the stationary points be $s_p(\varepsilon)$, then we have:
    \begin{equation}
      0 =
      E_\text{tot}(s_p(\varepsilon);\varepsilon) 
      = 
      \frac{1}{\Gamma_\varepsilon} 
      + R_\varepsilon(s_p(\varepsilon))
      .
    \label{eq:DefTotalForce}
    \end{equation}
    Depending on the situation, the dissipation strength or the resolvent may depend on the perturbation parameter $\varepsilon$.

    In each of the considered scenarios, we will find a relation between the approach of $s_p(\varepsilon)$ to its limit point and the behavior of the residual coefficient $r(\varepsilon) := R_\varepsilon(s_p(\varepsilon))/R'_\varepsilon(s_p(\varepsilon))$, which possibly is true in general.
    Namely:
    \begin{equation}
      \Landau{\Re[s_p(\varepsilon)]}
      =
      \Landau{ r(\varepsilon) }
      ,
    \label{eq:ResCoeffRel}
    \end{equation}
    as $\varepsilon\to0$. 
    Although this relation is not terribly important, it will be useful later to identify some asymptotically dominant terms.
    We don't know how to prove it in general.

    \subsection{Small dissipation limit}
      First, we consider the case $\Gamma \to 0$ (i.e. $\varepsilon = \Gamma$).
      This perturbation affects all stationary points.
      In this limit, the constant force becomes especially strong and must be balanced by an equally strong force from the point charges.
      The latter diverges at the points $-iE_l/\hbar$ and the stationary points must lie close to these poles.
      We thus make the ansatz $s_{p,l}(\Gamma) \AsymEq -iE_l/\hbar + c\Gamma$ with some constant $c$ to be determined.
      Plugging this into Eq.~\eqref{eq:DefTotalForce}, we find that all forces but the $l$-th charge's  are negligible
      \begin{equation}
        0 = \frac{1}{\Gamma} + \frac{p_l}{c\Gamma} + \Landau{1}
      \label{eq:}
      \end{equation}
      This gives:
      \begin{equation}
        s_{p,l}(\Gamma) \AsymEq = - \frac{i}{\hbar} E_l - p_l \Gamma
        .
      \label{eq:}
      \end{equation}
      The residual coefficients are given by:
      \begin{equation}
        r_l(\Gamma)
        =
        \frac{R(s_{p,l}(\Gamma))}{R'(s_{p,l}(\Gamma))}
        \AsymEq
        p_l \Gamma
        ,
      \label{eq:}
      \end{equation}
      showing Eq.~\eqref{eq:ResCoeffRel}.

    \subsection{Strong dissipation limit}
      \begin{figure}
        \includegraphics[width=0.99\columnwidth]{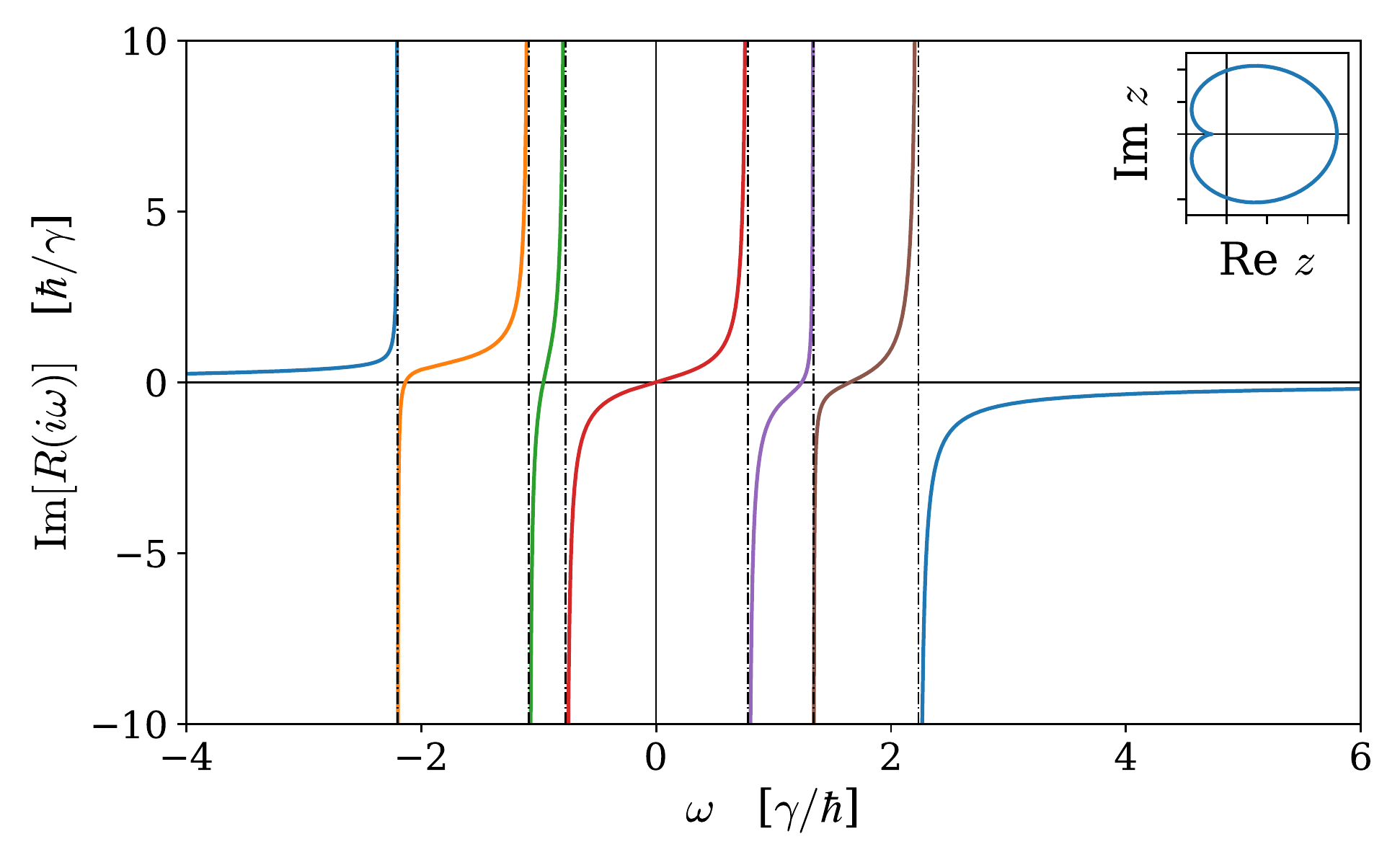}
        \caption{
          The resolvent $R(s)$, defined by Eq.~(6) maps the imaginary axis $w$ times to itself, where $w$ is given by the number of energy levels that have overlap with the decay state.
          Here we plot $R(s)$ for one realization of the disordered tight-binding model, Eq.~(15), with $L=6$ and disorder strength $\epsilon = \gamma$.
          The dash-dotted vertical lines are are the negative energy levels $-E_l/\hbar$.
          Each of the intervals $[-E_{l+1}/\hbar,-E_l/\hbar)$ and the outer segments get mapped once to the complete imaginary axis (different colors).
          The inset shows the curve $C(\Gamma R(\Bromwich))$ in the complex plane (i.e. Fig.~1 from the main text), whose winding number $w$ determines the mean decay time by Eq.~(2).
          \label{fig:Resolvent}
        }
      \end{figure}
      The next limit, we consider is opposite to before, namely when $\Gamma\to\infty$ ($\varepsilon = 1/\Gamma$).
      Again, all stationary points will be affected by the perturbation.
      From Eq.~\eqref{eq:DefTotalForce} we learn that the stationary points $s_p(\Gamma)$ must lie where the resolvent is very small.
      One of these areas is the far left half-plane, because $R(s) \AsymEq 1/s$, as $s\to\infty$.
      This determines one point, which we call ``the fast mode''.
      We thus make the ansatz: $s_{p,0}(\Gamma) \AsymEq - c \Gamma - i \omega_0$ and find:
      \begin{equation}
        0 
        = 
        \frac{1}{\Gamma}
        - \frac{1}{c\Gamma}\qty{ 1 - \frac{i}{\hbar} \Sum{l=1}{w} p_l [ \hbar \omega_0 - E_l ] }
        .
      \label{eq:}
      \end{equation}
      This identifies $c=1$, $\omega_0 = \sSum{l=1}{w} p_l E_l/\hbar = \ev*{\Ham}{\PsiDet}/\hbar$ and:
      \begin{equation}
        s_{p,0}(\Gamma)
        \AsymEq
        - \Gamma - \frac{i}{\hbar} \ev*{\Ham}{\PsiDet} 
        .
      \label{eq:}
      \end{equation}
      The remaining poles must lie close to the zeros $-i\omega_l$ of the resolvent $R(-i\omega_l)$.
      We make the ansatz $s_{p,l}(\varepsilon) \AsymEq - i \omega_l + c/\Gamma$ and find:
      \begin{equation}
        0 \AsymEq
        \frac{1}{\Gamma}
        + R(-i\omega_l) + R'(-i\omega_l) \frac{c}{\Gamma}
        .
      \label{eq:}
      \end{equation}
      This gives $c = - 1/R'(-i\omega_l)$ and
      \begin{equation}
        s_{p,l}(\varepsilon) 
        \AsymEq
        -i \omega_l - \frac{1}{\Gamma R'(-i\omega_l)}
        .
      \label{eq:}
      \end{equation}
      As depicted in Fig.~\ref{fig:Resolvent}, the resolvent has  $w-1$ zeros $-i\omega_l$.
      Each $\omega_l$ lies between two neighboring energy levels $E_l$ and $E_{l+1}$, see Fig.~\ref{fig:Resolvent}.
      Since the resolvent is monotonously increasing between such two levels, $\omega_l$ is particularly easy to find with standard root-finding algorithms like the Newton method.

      The residual coefficients are given by:
      \begin{equation}
        r_0(\Gamma)
        \AsymEq
        \Gamma
        \qc
        r_l(\Gamma) 
        \AsymEq
        \frac{1}{\Gamma R'(-i\omega_l) }
        .
      \label{eq:}
      \end{equation}
      Again Eq.~\eqref{eq:ResCoeffRel} is confirmed.

    \subsection{Small charges}
      Here, we consider the case of vanishing charges.
      Assume that of the total $w$ charges $p_l$, the first $W<w$ are small in the sense that $p_l(\varepsilon) \to 0$, for $1\le l \le W$, as $\varepsilon\to 0$.
      The corresponding energy levels $E_l$ and thus the positions of the charges remain unaffected.

      The total electrostatic force splits up into a foreground and a background part:
      \begin{equation}
        E_\text{BG}(s) = E_\text{tot}(s;0) = \frac{1}{\Gamma} + \sSum{l=W+1}{w} \frac{p_l}{s + \tfrac{i}{\hbar} E_l}
        .
      \label{eq:}
      \end{equation}
      The background force has $w-W$ stationary points that do not feel the perturbation at all.
      The remaining $W$ poles can be determined from a force balance between the background force and the force of a single charge.
      Clearly when $p_l(\varepsilon)\to0$ for $l\le W$, one stationary point will be pushed close to the point $-iE_l/\hbar$.
      Other small charges $p_{l'}(\varepsilon)$ with $l'\ne l$ and $l\le W$ are found to be negligible.
      We thus make the ansatz $s_{p,l} = -iE_l/\hbar + f(\varepsilon)$, and have:
      \begin{equation}
        0 
        = 
        \frac{p_l(\varepsilon)}{f(\varepsilon)}
        +
        E_\text{BG}(-i E_l/\hbar) + \landau{1}
        .
      \label{eq:}
      \end{equation}
      From here one determines $f(\varepsilon)$ and finds:
      \begin{equation}
        s_{p,l}(\varepsilon)
        \AsymEq
        - \frac{i}{\hbar} E_l - \frac{p_l(\varepsilon)}{E_\text{BG}(-\tfrac{i}{\hbar}E_l)}
        ,
      \label{eq:}
      \end{equation}
      as $\varepsilon\to0$ for $1\le l \le W$.
      The rate with which the charge vanishes is the same rate with which the stationary point approaches the charge.
      In contrast to the two previous scenarios, this rate can be different from charge to charge.
      In the long-time limit of $\PSI(t)$, only the fastest vanishing charge(s) are relevant.

      Finally, let us compute the residual coefficients of the affected stationary points:
      \begin{equation}
        r_l(\varepsilon)
        \AsymEq
        \frac{1}{\Gamma}
        \frac{
          p_l(\varepsilon)
        }{
          [E_\text{BG}(-\tfrac{i}{\hbar} E_l) ]^2
        }
        .
      \label{eq:}
      \end{equation}
      Yet another confirmation of Eq.~\eqref{eq:ResCoeffRel}.

      In the example of the two-level system, we have $p_- \AsymEq (\Omega/\delta)^2$, as $\Omega\to0$.
      Furthermore, we have $E_- = -\hbar \delta/2$ and $E_\text{BG}(-i E_- /\hbar) = 1/\Gamma - i /\delta$.
      Thus, we find the dominant pole and its residual coefficient:
      \begin{equation}
        s_{p,-} \AsymEq i\frac{\delta}{2} - \frac{\Omega^2\Gamma}{\delta^2 + \Gamma^2}
        \qc 
        r_- \AsymEq - \frac{\Omega^2}{(\delta - i \Gamma)^2}
        .
      \label{eq:}
      \end{equation}
      Eq.~\eqref{eq:HighMoments} gives the second moment of $T$.
      Taking only the dominant contribution into consideration gives:
      \begin{equation}
        \Var{T} 
        \AsymEq 
        \EA{T^2}
        \AsymEq
        \frac{4 \Gamma \abs{r}^2 }{(-2 \Re[s_p])^3}
        =
        \frac{\delta^2 + \Gamma^2}{2 \Gamma^2 \Omega^2}
        .
      \label{eq:}
      \end{equation}
      The same result is obtained from the leading order of Eq.~\eqref{eq:MomentsTwoLevel}.

    \subsection{Approximate degeneracy}
      The last considered situation is when $W$ charges are very close.
      We write $E_l(\varepsilon) = \bar{E} + \bar{E}_l\varepsilon$, for $1\le l\le W$, as $\varepsilon\to0$.
      Different from the last scenario, we here have to explicitly assume that all charges converge with the same rate to $\bar{E}$.
      Since the parametrization is arbitrary, we assume it linear in $\varepsilon$.
      The perturbation does not affect all stationary points, but only the first $W-1$ ones.
      The remaining $w-W+1$ stationary points are obtained from the background field, that treats the almost degenerate energy levels as one:
      \begin{equation}
        E_\text{BG}(s) := E_\text{tot}(s;0)
        =
        \frac{1}{\Gamma} 
        + \frac{\sSum{l=1}{W} p_l}{s + \tfrac{i}{\hbar} \bar{E}} 
        + \Sum{l=W+1}{w} \frac{p_l}{s + \tfrac{i}{\hbar} E_l}
        .
      \label{eq:}
      \end{equation}
      The foreground field consists of the close charges only.
      For $s\ApproxEq -i\bar{E}\hbar$, the foreground field blows up like $\sLandau{1/\varepsilon}$.
      We may thus define
      \begin{equation}
        \tilde{R}(s) 
        := 
        \Sum{l=1}{W}\frac{p_l}{s + \tfrac{i}{\hbar} \bar{E}_l}
        .
      \label{eq:}
      \end{equation}
      With the ansatz $s_{p,l}(\varepsilon) \AsymEq -i\bar{E}/\hbar - i\tilde{\omega}_l \varepsilon - \sigma_l\varepsilon^2$, we obtain the equation:
      \begin{equation}
        0 \AsymEq
        \frac{1}{\varepsilon} \tilde{R}(-i\tilde{\omega}_l)
        - \sigma_l\tilde{R}'(-i\tilde{\omega}_l)
        + E_\text{BG}(-\tfrac{i}{\hbar}\bar{E})
        .
      \label{eq:}
      \end{equation}
      For it to be satisfied, we must choose $-i\tilde{\omega}_l$ as the zeros of $\tilde{R}(s)$, and $\sigma_l = E_\text{BG}(-i\bar{E}/\hbar)/\tilde{R}'(-i\tilde{\omega}_l)$.
      In summary, the stationary points are:
      \begin{equation}
        s_{p,l}(\varepsilon)
        \AsymEq
        - \frac{E_\text{BG}(-\tfrac{i}{\hbar}\bar{E})}{\tilde{R}'(-i\tilde{\omega}_l)} \varepsilon^2
        - \frac{i}{\hbar} ( \bar{E} + \varepsilon \tilde{\omega}_l )
        .
      \label{eq:}
      \end{equation}
      Finally, we compute the residual coefficients:
      \begin{equation}
        r_l(\varepsilon)
        \AsymEq
        -
        \frac{\varepsilon^2}{\Gamma \tilde{R}'(-i\tilde{\omega}_l)} 
      \label{eq:}
      \end{equation}
      and confirm Eq.~\eqref{eq:ResCoeffRel} once again.

      This approximate-resonance scenario is exactly the relevant one for a degenerate energy level that is broken up by an external perturbation.
      Consider a Hamiltonian $\Ham_\varepsilon = \Ham_0 + \varepsilon \Ham_P$.
      $\Ham_0$ has the $g$-fold degenerate energy level $\bar{E}$ with eigenstates $\ket{E_m}$.
      This basis is already chosen such, that $\mel*{E_m}{\Ham_P}{E_{m'}} = \bar{E}_m \delta_{m,m'}$.
      The corresponding overlap for the unperturbed energy level is $p = \sSum{m=1}{g} p_m = \sSum{m=1}{g} \abs*{\ip*{\PsiDet}{E_m}}^2$.
      Under the perturbation $\Ham_P$, the energy level $\bar{E}$ will split up into $\bar{E} + \varepsilon \bar{E}_m$.
      The overlaps will also split, but in general none of them will vanish as $\varepsilon\to0$.
      If one of the overlaps is zero without the perturbation it will remain zero also under the perturbation.
      We stress that a perturbation of strength $\sLandau{\Delta E}$ yields slow decay rates of order $\sLandau{\Delta E^2}$.

  \section{Normalization of the dissipation time distribution}
  \label{app:TDP}
    In this section, we discuss the normalization $\TDP = \sInt{0}{\infty}{t} F(t)$ of the dissipation time distribution.
    This is done in two limits, $\Gamma\to0$, and $\Gamma\to\infty$, using the results of the last section.

    From Eq.~\eqref{eq:HighMoments} with $m=0$ we find in the small dissipation limit, using above results for the poles and residual coefficients:
    \begin{equation}
      \TDP
      =
      \Int{0}{\infty}{t} F(t)
      =
      1 
      +
      \Sum{l\ne l'=1}{w}
      \frac{2 \Gamma p_l p_{l'}}{
        (p_l + p_{l'})\Gamma + \frac{i}{\hbar}(E_l - E_{l'})
      }
      .
    \label{eq:TDPSmallDissLimit}
    \end{equation}
    Or in short $\TDP = 1 + \sLandau{\Gamma}$, as $\Gamma\to0$.

    In the strong dissipation limit, we find that the fast mode alone carries the normalization, as any terms involving $\omega_l$, $l\ge1$, give a subdominant contribution only:
    \begin{align}
      \TDP \AsymEq & \nonumber
      1
      +
      \Re{ \Sum{l=1}{w-1} \frac{
        4 \lambda_l 
      }{
        \Gamma^2 + \lambda_l + \frac{i}{\hbar} \Gamma (\omega_l - \omega_0)]
      } }
      \\ & 
      + 
      \Sum{l,l'=1}{w-1}
      \frac{
        2 \lambda_l \lambda_{l'} 
      }{ 
        (\lambda_l + \lambda_{l'})\Gamma^2  
        + \frac{i}{\hbar}\Gamma^3 (\omega_l - \omega_{l'})
      }
      .
    \label{eq:TDPLargeDissLimit}
    \end{align}
    In short $\TDP = 1 + \sLandau{\Gamma^{-2}}$ as $\Gamma\to\infty$.

    Eqs.~\eqref{eq:TDPSmallDissLimit} and \eqref{eq:TDPLargeDissLimit}, together with overwhelming numerical evidence, to be presented in Ref.~\cite{Thiel2020b}, convince us $\TDP=1$ holds in general.
    In terms of the poles and residual coefficients that implies the following sum rule:
    \begin{equation}
      \TDP
      =
      \Sum{l,l'=1}{w}
      \frac{r_l r_l^*}{(- s_{p,l} - s_{p,l'}^*)}
    \label{eq:}
    \end{equation}
    that we sadly can not prove in generality.

    Exceptions to the rule $\TDP =1$ occur when $\ket{\PsiDet}$ overlaps with the continuous spectrum of the Hamiltonian, or when another initial condition is used.

  \section{Finite observation time}
  \label{app:FinObs}
    In this section, we investigate the influence of large outliers in the dissipation time statistics.
    We start from Eq.~\eqref{eq:TimeDomainPSI}, which gives the dissipation time distribution in terms of the stationary points $s_{p,l}$ and the residual coefficients $r_l$ via $F(t) = 2\Gamma \abs{\PSI(t)}^2$:
    \begin{equation}
      F(t)
      =
      \frac{2}{\Gamma} \Sum{l,l'=1}{w} r_l r_{l'}^* e^{t(s_{p,l} + s_{p,l'}^*)}
      .
    \label{eq:PDFFromPoles}
    \end{equation}
    An experimenter, that does not wait infinitely long to record a dissipation event, but rather some finite time $\Theta$, does not see $F(t)$.
    When he records a histogram of dissipation times and removes from it all experiment runs with $T>\Theta$, he measures the conditional distribution $F_c(t) = F(t) / \sInt{0}{\Theta}{t} F(t)$.
    Computing the mean dissipation time from this distribution gives:
    \begin{equation}
      \EA{T_c(\Theta)}
      =
      \frac{
        \sInt{0}{\Theta}{t} t F(t) 
      }{
        \sInt{0}{\Theta}{t} F(t) 
      }
      ,
    \label{eq:}
    \end{equation}
    which is rewritten using Eq.~\eqref{eq:PDFFromPoles}:
    \begin{equation}
      \EA{T_c(\Theta)}
      =
      \frac{
        \frac{2}{\Gamma}
        \sSum{l,l'=1}{w} \frac{
          r_l r_{l'}^*
          \qty{ 1 - [1 - \Theta(s_{p,l} + s_{p,l'}^*) ]e^{\Theta(s_{p,l} + s_{p,l'}^*)} }
        }{
          (-s_{p,l} - s_{p,l'}^*)^2
        }
      }{
        \frac{2}{\Gamma}
        \sSum{l,l'=1}{w} \frac{
          r_l r_{l'}^*
        \qty[ 1 - e^{\Theta(s_{p,l} + s_{p,l'}^*)} ]
        }{
          (-s_{p,l} - s_{p,l'}^*)
        }
      }
      .
    \label{eq:CondMeanFull}
    \end{equation}
    As $\Theta$ becomes very large, the expressions in the braces tend to unity, because $\Re[s_{p,l}] < 0$.
    The denominator then converges to the normalization of $F(t)$ -- one -- and the numerator becomes the unconditioned dissipation time $\EA{T}$.
    The effect of finite observation times $\Theta < \infty$ is encoded in this expression.

    Let us consider a perturbation of strength $\varepsilon$ that moves the first $W<w$ poles very close to the imaginary axis, but does not significantly affect the other poles.
    That means, we assume that $\Re[s_{p,l}] = - \varepsilon \bar{s}_l$, for $l=1,\hdots,W$ where $\varepsilon$ is small, but $\Re[s_{p,l}] = \sLandau{1}$, for $l>W$.
    In the previous sections, we discussed several scenarios, where this happens.
    There, we also saw that the corresponding residual coefficients vanish with the same rate as $\Re[s_{p,l}]$, see Eq.~\eqref{eq:ResCoeffRel}.
    Therefore, we also take $r_l = \varepsilon \bar{r_l}$, for $l=1,\hdots,W$.
    Now, the leading order terms in Eq.~\eqref{eq:CondMeanFull} are collected.
    In the denominator, all summands with $l,l'\le W$ are $\sLandau{\varepsilon}$ and can be neglected.
    In the numerator this holds for all terms with $l,l'\le W$ and $l\ne l'$.
    We obtain:
    \begin{widetext}
      \begin{equation}
        \EA{T_c(\Theta)}
        \AsymEq
        \frac{
          \frac{2}{\Gamma}
          \sSum{l,l'=W+1}{w} \frac{
            r_l r_{l'}^*
            \qty{ 1 - [1 - \Theta(s_{p,l} + s_{p,l'}^*) ]e^{\Theta(s_{p,l} + s_{p,l'}^*)} }
          }{
            (-s_{p,l} - s_{p,l'}^*)^2
          }
        }{
          \frac{2}{\Gamma}
          \sSum{l,l'=W+1}{w} \frac{
            r_l r_{l'}^*
          \qty[ 1 - e^{\Theta(s_{p,l} + s_{p,l'}^*)} ]
          }{
            (-s_{p,l} - s_{p,l'}^*)
          }
        }
        +
        \frac{
          \frac{2}{\Gamma}
          \sSum{l=1}{W} 
          \frac{\abs{\bar{r_l}}^2}{4 \bar{s}_l^2}
          \qty{ 1 - [1 + 2 \bar{s}_l \varepsilon \Theta ] e^{-2\bar{s}_l \varepsilon\Theta} }
        }{
          \frac{2}{\Gamma}
          \sSum{l,l'=W+1}{w} \frac{
            r_l r_{l'}^*
          \qty[ 1 - e^{\Theta(s_{p,l} + s_{p,l'}^*)} ]
          }{
            (-s_{p,l} - s_{p,l'}^*)
          }
        }
        .
      \label{eq:CondMeanExpanded}
      \end{equation}
    \end{widetext}
    Clearly, the first term corresponds to the unperturbed mean dissipation time.
    Thus, when $\Theta$ is already so large that it fully resolves the unperturbed dynamics, we can replace the first term with $\EA{T}_{\varepsilon=0} = (w-W)/(2\Gamma)$
    The same can be said about the denominator of the second term, which can be replaced with one, making a mistake of order $\sLandau{\varepsilon}$.
    The numerator of the second term, however, is not a function of $\Theta$ or $\varepsilon$ alone, but rather of $\varepsilon\Theta$.
    Its argument can be small or large or anything in between.
    It thus describes a scaling function, that we call $g(x)$:
    \begin{equation}
      g(x) 
      := 
      \frac{4}{W}
      \Sum{l=1}{W}
      \frac{\abs{\bar{r_l}}^2}{4 \bar{s}_l^2}
      \qty{ 1 - [1 + 2 \bar{s}_l x ] e^{-2\bar{s}_l x} }
      .
    \label{eq:}
    \end{equation}
    We thus find:
    \begin{equation}
      \EA{T_c(\Theta)}
      =
      \frac{w-W}{2\Gamma} 
      + \frac{W}{2\Gamma} g(\varepsilon\Theta)
      .
    \label{eq:}
    \end{equation}
    From the definition of $g(x)$, we find that $g(x\to0) = 0$ and $g(x\to\infty) = 1+\sLandau{\varepsilon}$, such that it mediates a smooth crossover between the perturbed and unperturbed dissipation time as $\Theta$ increases.

    For the two-level system with small Rabi frequency we can use Eqs.~\eqref{eq:TwoLevelPoles} and \eqref{eq:TwoLevelResCoeffs} to write down $\sEA{T_c(\Theta)}$ explicitly.
    Expanding the poles $s_{p,\pm}$ and the residual coefficients $r_{\pm}$ for small $\Omega$ and plugging the result into Eq.~\eqref{eq:CondMeanExpanded} yields:
    \begin{align}
      \EA{T_c(\Theta)}
      \AsymEq & \nonumber
      \frac{1}{2\Gamma} \qty[
        1
        - \frac{2\Gamma\Theta e^{-2\Gamma\Theta}}{1 - e^{-2\Gamma\Theta}}
      ]
      + \\ & +
      \frac{1}{2\Gamma}
      \frac{
        1 - \qty[
          1 + \frac{2\Gamma\Omega^2\Theta}{\Gamma^2 + \delta^2}
        ] e^{-\frac{2\Gamma\Omega^2\Theta}{\Gamma^2 + \delta^2}}
      }{1 - e^{-2\Gamma\Theta}}
      .
    \label{eq:TwoLevelCondMeanExpanded}
    \end{align}
    When $\Theta \gg 1/(2\Gamma)$, we can neglect all exponentials $e^{-2\Gamma\Theta}$.
    This way only functions of $\Omega^2\Theta$ survive, and we can identify the scaling function $g(x)$: 
    \begin{equation}
      g(x)
      =
      1 - \qty[
        1 + \frac{2\Gamma x}{\Gamma^2 + \delta^2}
      ] e^{-\frac{2\Gamma x}{\Gamma^2 + \delta^2}}
      ,
    \label{eq:TwoLevelScaling}
    \end{equation}
    and find
    \begin{equation}
      \EA{T_c(\Theta)} = \frac{1}{2\Gamma} \qty[ 1 + g(\Omega^2\Theta) ]
      .
    \label{eq:}
    \end{equation}
    This is Eq.~\eqref{eq:TwoLevelCondMean} of the main text and also the solid line plotted in Fig.~\ref{fig:TwoLevel}(B).
    The symbols in that figure are the full result from Eq.~\eqref{eq:CondMeanFull}.

\section{General preparation states}
\label{app:Arr}
  \begin{figure}
    \includegraphics[width=0.99\columnwidth]{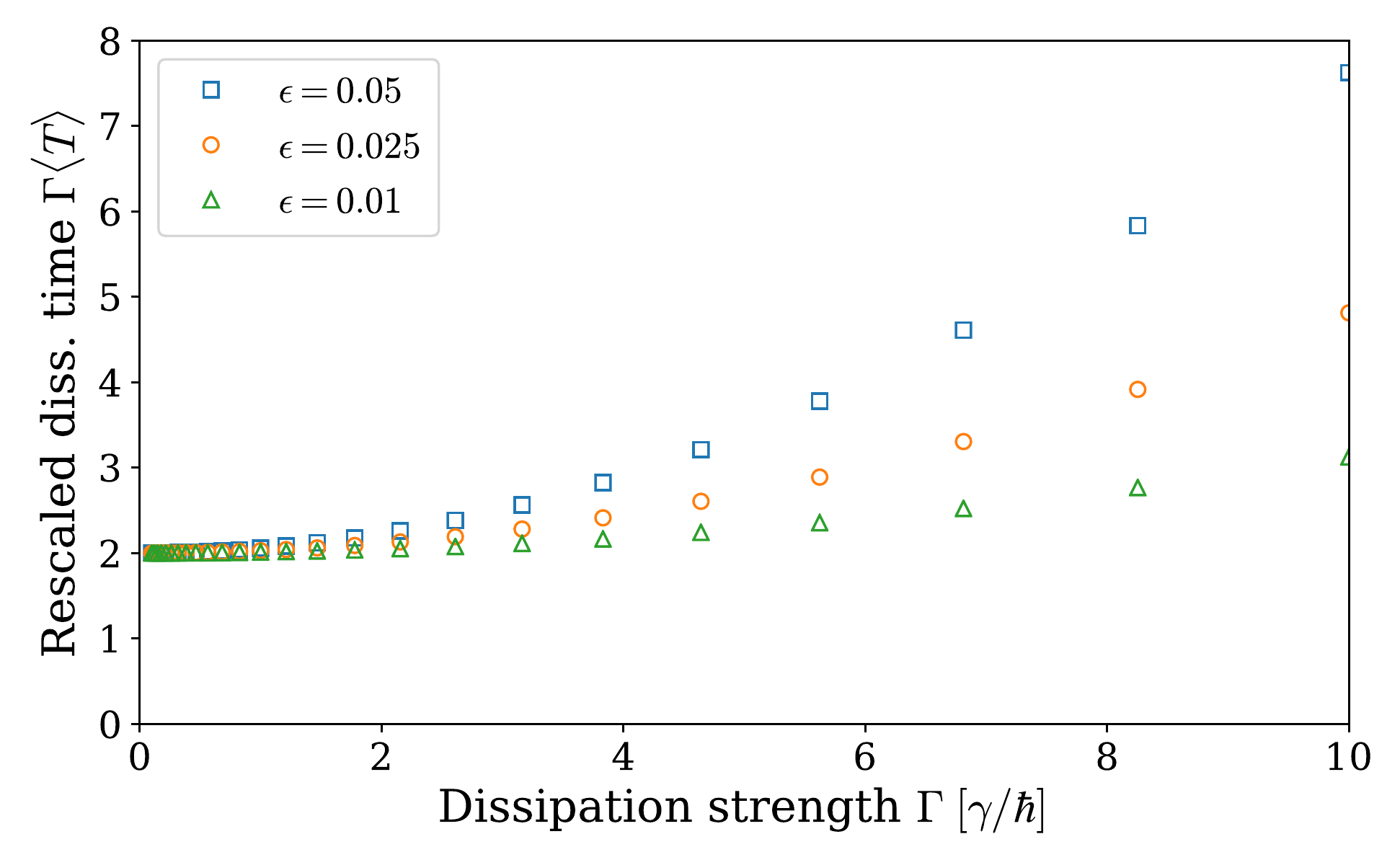}
    \caption{
      Initial state different from dissipation state.
      Here, for the tight-binding model, Eq.~\eqref{eq:TBHam}, with $L=6$ and $\epsilon=0$.
      Initial state is given by Eq.~\eqref{eq:ArrIn}, and $\delta = 1 - \abs*{\ip*{\PsiIn}{\PsiDet}}^2$ quantifies the (large) overlap between initial and dissipation state.
      Discrepancies from the quantized result ($w=4$, solid line) are proportional to $\delta$.
      They vanish as $\Gamma$ goes to zero, but are significant for larger $\Gamma$.
      \label{fig:Arr}
    }
  \end{figure}
  Our main result, the strict quantization of the mean dissipation time, crucially depends on the choice of the initial state $\ket{\PsiIn} = \ket{\psi(t=0)} = \ket{\PsiDet}$.
  In this section, we shortly present he system's behavior for different initial states, that are close to the dissipation state.
  In particular, we consider here the tight-binding model on the ring with six sites without disorder, i.e. Eq.~\eqref{eq:TBHam} with $L=6$ and $\epsilon = 0$.
  The dissipation channel is localized on the $L$th site $\ket{\PsiDet} = \ket{L} = \ket{6}$ and the initial state is chosen to be:
  \begin{equation}
    \ket{\PsiIn} = \sqrt{1 - \delta} \ket{6} + \sqrt{\delta} \ket{3}
    ,
  \label{eq:ArrIn}
  \end{equation}
  such that the overlap between the initial and detection state is given by $\abs*{\ip*{\PsiIn}{\PsiDet}}^2 = 1 -  \delta$, where $\delta$ is chosen small.
  The rescaled mean dissipation time for these parameters is depicted in Fig.~\ref{fig:Arr} as a function of $\Gamma$ for three values of $\delta$.
  As we will explore in more generality in another publication \cite{Thiel2020b}, the mean dissipation time is approximately quantized as long as either $\delta$ or $\Gamma$ is very small.
  For general initial states one finds that $\EA{T} \AsymEq \Gamma$, which explains the seemingly parabolic shape of the curves in Fig.~\ref{fig:Arr}.
  There are, however, more pitfalls, because dissipation is not necessarily ensured for general initial states.
  This means that possibly $\TDP = \sInt{0}{\infty}{t} F(t) < 1$, such that $\EA{T} = \sInt{0}{\infty}{t} t F(t) / \TDP$ can only be interpreted in a conditional sense.

\section{Multiple dissipation channels}
\label{app:MultiDiss}
  In this section, we push the analogy between the non-Hermitian Schr\"odinger equation and unitary evolution disturbed by projective stroboscopic measurements.
  Consider a system with Hamiltonian $\Ham = \sSum{l}{} \hat{P}_l E_l$, where $\hat{P}_l = \sSum{m=1}{g_l} \dyad{E_{l,m}}$.
  Every $\tau$ time units a projective measurement tests whether it resides in some detection space with orthogonal projector $\Detect := \sSum{j=1}{d} \dyad*{\PsiDet^j}$, with $\ip*{\PsiDet^j}{\PsiDet^{j'}} = \delta_{j,j'}$.
  The random time $T = n\tau$ of the first successful detection attempts is the first detection time.
  Ref.~\cite{Bourgain2014a} proved that the mean first detection time is quantized by $\sEA{T} = (w / d)\tau$, provided that the system is prepared in a completely mixed state over the detection subspace, i.e. $\hat{\rho}_\text{in} = \Detect/d$.
  Our follow-up manuscript \cite{Thiel2020b} investigates the equivalence between the first detection time and the mean dissipation time in depth for $d= \mathrm{rank} \Detect = 1$.
  From this equivalence and the quantization for higher-dimensional subspaces in the quantum first detection problem, we conjecture that the mean dissipation time for the non-Hermitian Schr\"odinger equation is also quantized for more than one dissipation channel.

  More precisely, consider the non-Hermitian Schr\"odinger equation:
  \begin{equation}
    i \hbar \pdv{t} \ket{\psi(t)} = \Ham \ket{\psi(t)} - i \hbar \Gamma \Detect \ket{\psi(t)}
    ,
  \label{eq:}
  \end{equation}
  where $\Detect$ as before.
  The system is prepared in the mixed initial state $\hat{\rho}_\text{in} = \Detect/d$, that is to say, with probability $1/d$ one chooses the pure initial state $\ket{\psi(t=0)} = \ket*{\PsiDet^j}$.
  Let $\hat{U}(t) := e^{-i(t/\hbar) \Ham - t \Gamma \Detect}$, then the momentary probability of absorption is clearly given by $F(t) = 2 \Gamma \Trace[\Detect \hat{U}(t) \hat{\rho}_\text{in} [\hat{U}(t)]^\dagger ]$.
  Translating the stroboscopic result, we conjecture a rational quantization: 
  \begin{equation}
    \EA{T} = \frac{w}{2 d \Gamma}
    ,
  \label{eq:}
  \end{equation}
  where $d = \mathrm{rank} \Detect$ is the number of different dissipation channels and the winding number is $w = \sSum{l}{} \mathrm{rank} [ \Detect \hat{P}_l ]$, see \cite[Thm. 3.2]{Bourgain2014a}.
  Eq.~(2) of the main text is a special case of this equation with $d=1$.
  The new definition of the winding number implies that each degenerate energy level may contribute to a variable extend, depending on the symmetry properties of the dissipation states.

\begin{figure}
    \includegraphics[width=0.99\columnwidth]{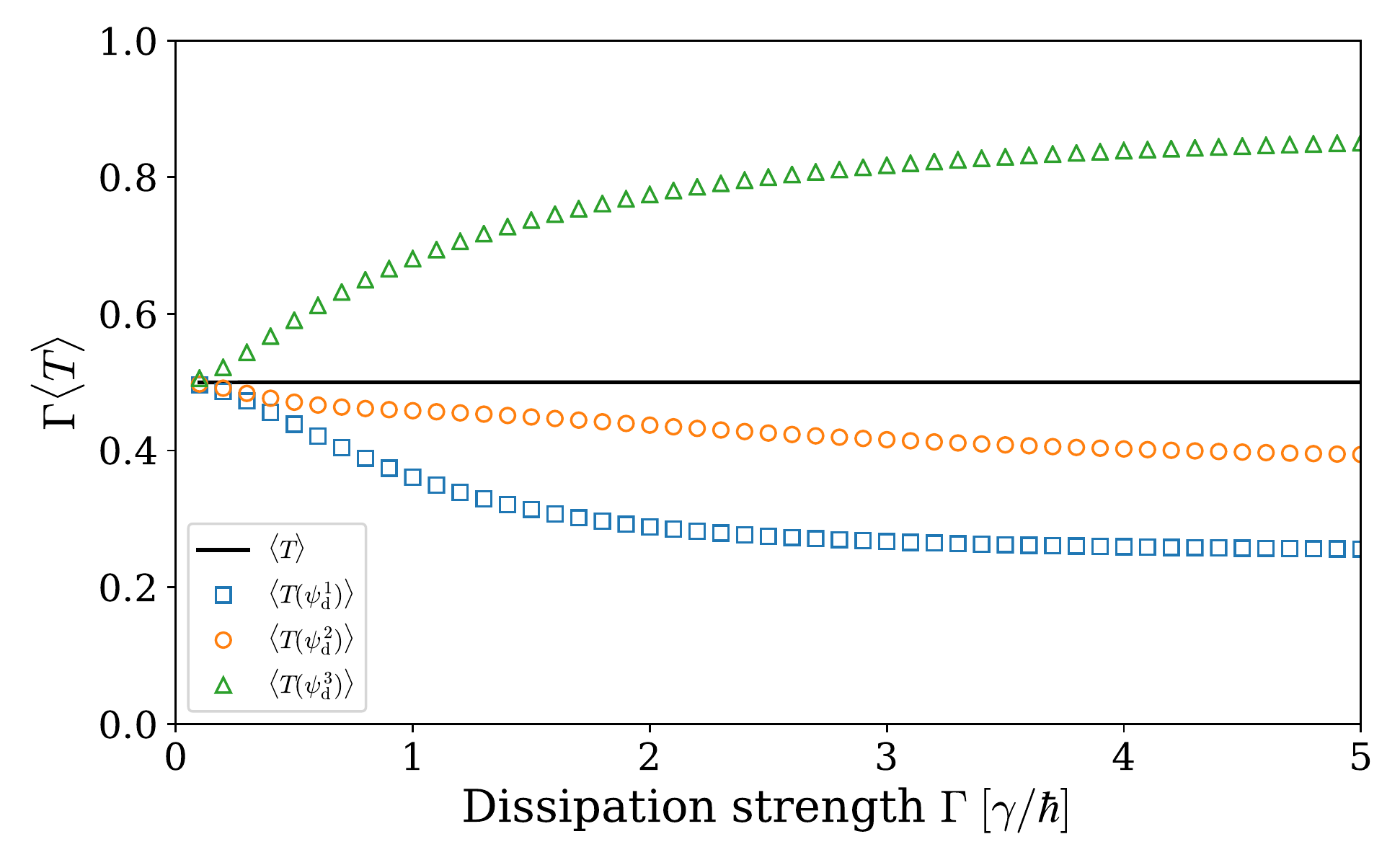}
    \caption{
      Rescaled mean dissipation time for a line with six sites and three dissipation channels $\ket*{\PsiDet^1} = \ket{1}$, $\ket*{\PsiDet^2} = \ket{2}$, and $\ket*{\PsiDet^3} = \ket{4}$.
      The straight line depicts the quantized mean for the mixed initial condition.
      The symbols depict the mean dissipation time when the system is prepared in each individual dissipation state (although all three channels are dissipative).
      Each individual curve is not quantized, but their average is.
      Here, $w=3$ and $d=3$.
      \label{fig:MultiDiss}
    }
  \end{figure}

  We tested this hypothesis in a simple line Hamiltonian with six sites, i.e. Eq.~\eqref{eq:TBHam} with $L=6$, $\epsilon=0$ and reflecting boundary conditions.
  In Fig.~\ref{fig:MultiDiss}, we depicted the rescaled mean dissipation time $\Gamma \EA{T}$ for the case when $\ket*{\PsiDet^1} = \ket{1}$, $\ket*{\PsiDet^2} = \ket{2}$, and $\ket*{\PsiDet^3} = \ket{4}$.
  Clearly, the above defined dissipation time is an average of three constituents $\sEA{T(\PsiDet^j)}$.
  These are the dissipation times for a system prepared in the pure state $\ket*{\PsiDet^j}$.
  Interestingly, these individual dissipation times are not quantized and exhibit some non-trivial dependence on $\Gamma$.
  Their average, however, is quantized, as shown in Fig.~\ref{fig:MultiDiss}.


\begin{thebibliography}{49}
\expandafter\ifx\csname natexlab\endcsname\relax\def\natexlab#1{#1}\fi
\expandafter\ifx\csname bibnamefont\endcsname\relax
  \def\bibnamefont#1{#1}\fi
\expandafter\ifx\csname bibfnamefont\endcsname\relax
  \def\bibfnamefont#1{#1}\fi
\expandafter\ifx\csname citenamefont\endcsname\relax
  \def\citenamefont#1{#1}\fi
\expandafter\ifx\csname url\endcsname\relax
  \def\url#1{\texttt{#1}}\fi
\expandafter\ifx\csname urlprefix\endcsname\relax\def\urlprefix{URL }\fi
\providecommand{\bibinfo}[2]{#2}
\providecommand{\eprint}[2][]{\url{#2}}

\bibitem[{\citenamefont{Berry}(1984)}]{Berry1984a}
\bibinfo{author}{\bibfnamefont{M.~V.} \bibnamefont{Berry}},
  \bibinfo{journal}{Procedings of the Royal Society A}  (\bibinfo{year}{1984}).

\bibitem[{\citenamefont{Zak}(1989)}]{Zak1989a}
\bibinfo{author}{\bibfnamefont{J.}~\bibnamefont{Zak}},
  \bibinfo{journal}{Physical Review Letters} \textbf{\bibinfo{volume}{62}},
  \bibinfo{pages}{2747} (\bibinfo{year}{1989}),
  \urlprefix\url{https://link.aps.org/doi/10.1103/PhysRevLett.62.2747}.

\bibitem[{\citenamefont{Gamow}(1928)}]{Gamow1928a}
\bibinfo{author}{\bibfnamefont{G.}~\bibnamefont{Gamow}},
  \bibinfo{journal}{Zeitschrift für Physik} \textbf{\bibinfo{volume}{51}},
  \bibinfo{pages}{204} (\bibinfo{year}{1928}), ISSN \bibinfo{issn}{0044-3328},
  \urlprefix\url{https://doi.org/10.1007/BF01343196}.

\bibitem[{\citenamefont{Feshbach}(1958)}]{Feshbach1958a}
\bibinfo{author}{\bibfnamefont{H.}~\bibnamefont{Feshbach}},
  \bibinfo{journal}{Annals of Physics} \textbf{\bibinfo{volume}{5}},
  \bibinfo{pages}{357 } (\bibinfo{year}{1958}), ISSN \bibinfo{issn}{0003-4916},
  \urlprefix\url{http://www.sciencedirect.com/science/article/pii/0003491658900071}.

\bibitem[{\citenamefont{Ho}(1983)}]{Ho1983a}
\bibinfo{author}{\bibfnamefont{Y.~K.} \bibnamefont{Ho}},
  \bibinfo{journal}{Physics Reports} \textbf{\bibinfo{volume}{99}},
  \bibinfo{pages}{1} (\bibinfo{year}{1983}).

\bibitem[{\citenamefont{Buchleitner et~al.}(1994)\citenamefont{Buchleitner,
  Gr{\'e}maud, and Delande}}]{Buchleitner1994a}
\bibinfo{author}{\bibfnamefont{A.}~\bibnamefont{Buchleitner}},
  \bibinfo{author}{\bibfnamefont{B.}~\bibnamefont{Gr{\'e}maud}},
  \bibnamefont{and} \bibinfo{author}{\bibfnamefont{D.}~\bibnamefont{Delande}},
  \bibinfo{journal}{J Phys B: At. Mol. Opt. Phys.}
  \textbf{\bibinfo{volume}{27}}, \bibinfo{pages}{2663} (\bibinfo{year}{1994}).

\bibitem[{\citenamefont{Bender}(2007)}]{Bender2007a}
\bibinfo{author}{\bibfnamefont{C.~M.} \bibnamefont{Bender}},
  \bibinfo{journal}{Reports on Progress in Physics}
  \textbf{\bibinfo{volume}{70}}, \bibinfo{pages}{947} (\bibinfo{year}{2007}),
  \urlprefix\url{https://doi.org/10.1088%2F0034-4885%2F70%2F6%2Fr03}.

\bibitem[{\citenamefont{Moiseyev}(2011)}]{Moiseyev2011a}
\bibinfo{author}{\bibfnamefont{N.}~\bibnamefont{Moiseyev}},
  \emph{\bibinfo{title}{Non-Hermitian Quantum Mechanics}}
  (\bibinfo{publisher}{Campbridge University Press, Cambridge},
  \bibinfo{year}{2011}), ISBN \bibinfo{isbn}{978-0521889728}.

\bibitem[{\citenamefont{Philip et~al.}(2018)\citenamefont{Philip, Hirsbrunner,
  and Gilbert}}]{Philip2018a}
\bibinfo{author}{\bibfnamefont{T.~M.} \bibnamefont{Philip}},
  \bibinfo{author}{\bibfnamefont{M.~R.} \bibnamefont{Hirsbrunner}},
  \bibnamefont{and} \bibinfo{author}{\bibfnamefont{M.~J.}
  \bibnamefont{Gilbert}}, \bibinfo{journal}{Physical Review B}
  \textbf{\bibinfo{volume}{98}}, \bibinfo{pages}{155430}
  (\bibinfo{year}{2018}),
  \urlprefix\url{https://link.aps.org/doi/10.1103/PhysRevB.98.155430}.

\bibitem[{\citenamefont{Chen and Zhai}(2018)}]{Chen2018a}
\bibinfo{author}{\bibfnamefont{Y.}~\bibnamefont{Chen}} \bibnamefont{and}
  \bibinfo{author}{\bibfnamefont{H.}~\bibnamefont{Zhai}},
  \bibinfo{journal}{Physical Review B} \textbf{\bibinfo{volume}{98}},
  \bibinfo{pages}{245130} (\bibinfo{year}{2018}),
  \urlprefix\url{https://link.aps.org/doi/10.1103/PhysRevB.98.245130}.

\bibitem[{\citenamefont{Rudner and Levitov}(2009)}]{Rudner2009a}
\bibinfo{author}{\bibfnamefont{M.~S.} \bibnamefont{Rudner}} \bibnamefont{and}
  \bibinfo{author}{\bibfnamefont{L.~S.} \bibnamefont{Levitov}},
  \bibinfo{journal}{Physical Review Letters} \textbf{\bibinfo{volume}{102}},
  \bibinfo{pages}{065703} (\bibinfo{year}{2009}), ISSN
  \bibinfo{issn}{1079-7114},
  \urlprefix\url{http://dx.doi.org/10.1103/PhysRevLett.102.065703}.

\bibitem[{\citenamefont{Konotop et~al.}(2016)\citenamefont{Konotop, Yang, and
  Zezyulin}}]{Konotop2016a}
\bibinfo{author}{\bibfnamefont{V.~V.} \bibnamefont{Konotop}},
  \bibinfo{author}{\bibfnamefont{J.}~\bibnamefont{Yang}}, \bibnamefont{and}
  \bibinfo{author}{\bibfnamefont{D.~A.} \bibnamefont{Zezyulin}},
  \bibinfo{journal}{Reviews of Modern Physics} \textbf{\bibinfo{volume}{88}},
  \bibinfo{pages}{035002} (\bibinfo{year}{2016}),
  \urlprefix\url{https://link.aps.org/doi/10.1103/RevModPhys.88.035002}.

\bibitem[{\citenamefont{El-Ganainy et~al.}(2018)\citenamefont{El-Ganainy,
  Makris, Khajavikhan, Musslimani, Rotter, and
  Christodoulides}}]{El-Ganainy2018a}
\bibinfo{author}{\bibfnamefont{R.}~\bibnamefont{El-Ganainy}},
  \bibinfo{author}{\bibfnamefont{K.~G.} \bibnamefont{Makris}},
  \bibinfo{author}{\bibfnamefont{M.}~\bibnamefont{Khajavikhan}},
  \bibinfo{author}{\bibfnamefont{Z.~H.} \bibnamefont{Musslimani}},
  \bibinfo{author}{\bibfnamefont{S.}~\bibnamefont{Rotter}}, \bibnamefont{and}
  \bibinfo{author}{\bibfnamefont{D.~N.} \bibnamefont{Christodoulides}},
  \bibinfo{journal}{Nature Physics} \textbf{\bibinfo{volume}{14}},
  \bibinfo{pages}{11} (\bibinfo{year}{2018}), ISSN \bibinfo{issn}{1745-2481},
  \urlprefix\url{https://doi.org/10.1038/nphys4323}.

\bibitem[{\citenamefont{Gong et~al.}(2018)\citenamefont{Gong, Ashida, Kawabata,
  Takasan, Higashikawa, and Ueda}}]{Gong2018a}
\bibinfo{author}{\bibfnamefont{Z.}~\bibnamefont{Gong}},
  \bibinfo{author}{\bibfnamefont{Y.}~\bibnamefont{Ashida}},
  \bibinfo{author}{\bibfnamefont{K.}~\bibnamefont{Kawabata}},
  \bibinfo{author}{\bibfnamefont{K.}~\bibnamefont{Takasan}},
  \bibinfo{author}{\bibfnamefont{S.}~\bibnamefont{Higashikawa}},
  \bibnamefont{and} \bibinfo{author}{\bibfnamefont{M.}~\bibnamefont{Ueda}},
  \bibinfo{journal}{Physical Review X} \textbf{\bibinfo{volume}{8}},
  \bibinfo{pages}{031079} (\bibinfo{year}{2018}),
  \urlprefix\url{https://link.aps.org/doi/10.1103/PhysRevX.8.031079}.

\bibitem[{\citenamefont{Ghatak and Das}(2019)}]{Ghatak2019a}
\bibinfo{author}{\bibfnamefont{A.}~\bibnamefont{Ghatak}} \bibnamefont{and}
  \bibinfo{author}{\bibfnamefont{T.}~\bibnamefont{Das}},
  \bibinfo{journal}{Journal of Physics: Condensed Matter}
  \textbf{\bibinfo{volume}{31}}, \bibinfo{pages}{263001}
  (\bibinfo{year}{2019}),
  \urlprefix\url{https://doi.org/10.1088%2F1361-648x%2Fab11b3}.

\bibitem[{\citenamefont{Kawabata et~al.}(2019)\citenamefont{Kawabata, Shiozaki,
  Ueda, and Sato}}]{Kawabata2019a}
\bibinfo{author}{\bibfnamefont{K.}~\bibnamefont{Kawabata}},
  \bibinfo{author}{\bibfnamefont{K.}~\bibnamefont{Shiozaki}},
  \bibinfo{author}{\bibfnamefont{M.}~\bibnamefont{Ueda}}, \bibnamefont{and}
  \bibinfo{author}{\bibfnamefont{M.}~\bibnamefont{Sato}},
  \bibinfo{journal}{Physical Review X} \textbf{\bibinfo{volume}{9}},
  \bibinfo{pages}{041015} (\bibinfo{year}{2019}),
  \urlprefix\url{https://link.aps.org/doi/10.1103/PhysRevX.9.041015}.

\bibitem[{\citenamefont{Xu et~al.}(2017)\citenamefont{Xu, Wang, and
  Duan}}]{Xu2017a}
\bibinfo{author}{\bibfnamefont{Y.}~\bibnamefont{Xu}},
  \bibinfo{author}{\bibfnamefont{S.-T.} \bibnamefont{Wang}}, \bibnamefont{and}
  \bibinfo{author}{\bibfnamefont{L.-M.} \bibnamefont{Duan}},
  \bibinfo{journal}{Physical Review Letters} \textbf{\bibinfo{volume}{118}},
  \bibinfo{pages}{045701} (\bibinfo{year}{2017}),
  \urlprefix\url{https://link.aps.org/doi/10.1103/PhysRevLett.118.045701}.

\bibitem[{\citenamefont{Rivet et~al.}(2018)\citenamefont{Rivet, Brandstötter,
  Makris, Lissek, Rotter, and Fleury}}]{Rivet2018a}
\bibinfo{author}{\bibfnamefont{E.}~\bibnamefont{Rivet}},
  \bibinfo{author}{\bibfnamefont{A.}~\bibnamefont{Brandstötter}},
  \bibinfo{author}{\bibfnamefont{K.~G.} \bibnamefont{Makris}},
  \bibinfo{author}{\bibfnamefont{H.}~\bibnamefont{Lissek}},
  \bibinfo{author}{\bibfnamefont{S.}~\bibnamefont{Rotter}}, \bibnamefont{and}
  \bibinfo{author}{\bibfnamefont{R.}~\bibnamefont{Fleury}},
  \bibinfo{journal}{Nature Physics} \textbf{\bibinfo{volume}{14}},
  \bibinfo{pages}{942} (\bibinfo{year}{2018}), ISSN \bibinfo{issn}{1745-2481},
  \urlprefix\url{https://doi.org/10.1038/s41567-018-0188-7}.

\bibitem[{\citenamefont{Xiao et~al.}(2019)\citenamefont{Xiao, Wang, Zhan, Bian,
  Kawabata, Ueda, Yi, and Xue}}]{Xiao2019a}
\bibinfo{author}{\bibfnamefont{L.}~\bibnamefont{Xiao}},
  \bibinfo{author}{\bibfnamefont{K.}~\bibnamefont{Wang}},
  \bibinfo{author}{\bibfnamefont{X.}~\bibnamefont{Zhan}},
  \bibinfo{author}{\bibfnamefont{Z.}~\bibnamefont{Bian}},
  \bibinfo{author}{\bibfnamefont{K.}~\bibnamefont{Kawabata}},
  \bibinfo{author}{\bibfnamefont{M.}~\bibnamefont{Ueda}},
  \bibinfo{author}{\bibfnamefont{W.}~\bibnamefont{Yi}}, \bibnamefont{and}
  \bibinfo{author}{\bibfnamefont{P.}~\bibnamefont{Xue}},
  \bibinfo{journal}{Physical Review Letters} \textbf{\bibinfo{volume}{123}},
  \bibinfo{pages}{230401} (\bibinfo{year}{2019}),
  \urlprefix\url{https://link.aps.org/doi/10.1103/PhysRevLett.123.230401}.

\bibitem[{\citenamefont{Lapp et~al.}(2019)\citenamefont{Lapp, Ang'ong'a, An,
  and Gadway}}]{Lapp2019a}
\bibinfo{author}{\bibfnamefont{S.}~\bibnamefont{Lapp}},
  \bibinfo{author}{\bibfnamefont{J.}~\bibnamefont{Ang'ong'a}},
  \bibinfo{author}{\bibfnamefont{F.~A.} \bibnamefont{An}}, \bibnamefont{and}
  \bibinfo{author}{\bibfnamefont{B.}~\bibnamefont{Gadway}},
  \bibinfo{journal}{New Journal of Physics} \textbf{\bibinfo{volume}{21}},
  \bibinfo{pages}{045006} (\bibinfo{year}{2019}),
  \urlprefix\url{https://doi.org/10.1088%2F1367-2630%2Fab1147}.

\bibitem[{\citenamefont{Li et~al.}(2019)\citenamefont{Li, Harter, Liu, de~Melo,
  Joglekar, and Luo}}]{Li2019a}
\bibinfo{author}{\bibfnamefont{J.}~\bibnamefont{Li}},
  \bibinfo{author}{\bibfnamefont{A.~K.} \bibnamefont{Harter}},
  \bibinfo{author}{\bibfnamefont{J.}~\bibnamefont{Liu}},
  \bibinfo{author}{\bibfnamefont{L.}~\bibnamefont{de~Melo}},
  \bibinfo{author}{\bibfnamefont{Y.~N.} \bibnamefont{Joglekar}},
  \bibnamefont{and} \bibinfo{author}{\bibfnamefont{L.}~\bibnamefont{Luo}},
  \bibinfo{journal}{Nature Communications} \textbf{\bibinfo{volume}{10}},
  \bibinfo{pages}{855} (\bibinfo{year}{2019}), ISSN \bibinfo{issn}{2041-1723},
  \urlprefix\url{https://doi.org/10.1038/s41467-019-08596-1}.

\bibitem[{\citenamefont{Caruso et~al.}(2009)\citenamefont{Caruso, Chin, Datta,
  Huelga, and Plenio}}]{Caruso2009a}
\bibinfo{author}{\bibfnamefont{F.}~\bibnamefont{Caruso}},
  \bibinfo{author}{\bibfnamefont{A.~W.} \bibnamefont{Chin}},
  \bibinfo{author}{\bibfnamefont{A.}~\bibnamefont{Datta}},
  \bibinfo{author}{\bibfnamefont{S.~F.} \bibnamefont{Huelga}},
  \bibnamefont{and} \bibinfo{author}{\bibfnamefont{M.~B.}
  \bibnamefont{Plenio}}, \bibinfo{journal}{The Journal of Chemical Physics}
  \textbf{\bibinfo{volume}{131}}, \bibinfo{pages}{105106}
  (\bibinfo{year}{2009}),
  \eprint{https://aip.scitation.org/doi/pdf/10.1063/1.3223548}.

\bibitem[{\citenamefont{Agliari et~al.}(2010)\citenamefont{Agliari, M\"{u}lken,
  and Blumen}}]{Agliari2010a}
\bibinfo{author}{\bibfnamefont{E.}~\bibnamefont{Agliari}},
  \bibinfo{author}{\bibfnamefont{O.}~\bibnamefont{M\"{u}lken}},
  \bibnamefont{and} \bibinfo{author}{\bibfnamefont{A.}~\bibnamefont{Blumen}},
  \bibinfo{journal}{International Journal of Bifurcation and Chaos}
  \textbf{\bibinfo{volume}{20}}, \bibinfo{pages}{271} (\bibinfo{year}{2010}),
  \urlprefix\url{https://doi.org/10.1142/S0218127410025715}.

\bibitem[{\citenamefont{M{\"u}lken and Blumen}(2011)}]{Muelken2011a}
\bibinfo{author}{\bibfnamefont{O.}~\bibnamefont{M{\"u}lken}} \bibnamefont{and}
  \bibinfo{author}{\bibfnamefont{A.}~\bibnamefont{Blumen}},
  \bibinfo{journal}{Physics Reports} \textbf{\bibinfo{volume}{502}},
  \bibinfo{pages}{37} (\bibinfo{year}{2011}).

\bibitem[{\citenamefont{Krapivsky et~al.}(2014)\citenamefont{Krapivsky, Luck,
  and Mallick}}]{Krapivsky2014a}
\bibinfo{author}{\bibfnamefont{P.~L.} \bibnamefont{Krapivsky}},
  \bibinfo{author}{\bibfnamefont{J.~M.} \bibnamefont{Luck}}, \bibnamefont{and}
  \bibinfo{author}{\bibfnamefont{K.}~\bibnamefont{Mallick}},
  \bibinfo{journal}{Journal of Statistical Physics}
  \textbf{\bibinfo{volume}{154}}, \bibinfo{pages}{1430} (\bibinfo{year}{2014}),
  ISSN \bibinfo{issn}{1572-9613},
  \urlprefix\url{http://dx.doi.org/10.1007/s10955-014-0936-8}.

\bibitem[{\citenamefont{Novo et~al.}(2015)\citenamefont{Novo, Chakraborty,
  Mohseni, Neven, and Omar}}]{Novo2015a}
\bibinfo{author}{\bibfnamefont{L.}~\bibnamefont{Novo}},
  \bibinfo{author}{\bibfnamefont{S.}~\bibnamefont{Chakraborty}},
  \bibinfo{author}{\bibfnamefont{M.}~\bibnamefont{Mohseni}},
  \bibinfo{author}{\bibfnamefont{H.}~\bibnamefont{Neven}}, \bibnamefont{and}
  \bibinfo{author}{\bibfnamefont{Y.}~\bibnamefont{Omar}},
  \bibinfo{journal}{Scientific Reports} \textbf{\bibinfo{volume}{5}},
  \bibinfo{pages}{13304} (\bibinfo{year}{2015}),
  \urlprefix\url{www.nature.com/articles/srep13304}.

\bibitem[{\citenamefont{Giusteri et~al.}(2015)\citenamefont{Giusteri,
  Mattiotti, and Celardo}}]{Giusteri2015a}
\bibinfo{author}{\bibfnamefont{G.~G.} \bibnamefont{Giusteri}},
  \bibinfo{author}{\bibfnamefont{F.}~\bibnamefont{Mattiotti}},
  \bibnamefont{and} \bibinfo{author}{\bibfnamefont{G.~L.}
  \bibnamefont{Celardo}}, \bibinfo{journal}{Physical Review B}
  \textbf{\bibinfo{volume}{91}}, \bibinfo{pages}{094301}
  (\bibinfo{year}{2015}).

\bibitem[{\citenamefont{Plenio and Knight}(1998)}]{Plenio1998a}
\bibinfo{author}{\bibfnamefont{M.~B.} \bibnamefont{Plenio}} \bibnamefont{and}
  \bibinfo{author}{\bibfnamefont{P.~L.} \bibnamefont{Knight}},
  \bibinfo{journal}{Reviews of Modern Physics} \textbf{\bibinfo{volume}{70}},
  \bibinfo{pages}{101} (\bibinfo{year}{1998}).

\bibitem[{\citenamefont{Gisin and Percival}(1992)}]{Gisin1992a}
\bibinfo{author}{\bibfnamefont{N.}~\bibnamefont{Gisin}} \bibnamefont{and}
  \bibinfo{author}{\bibfnamefont{I.~C.} \bibnamefont{Percival}},
  \bibinfo{journal}{Journal of Physics A: Mathematical and General}
  \textbf{\bibinfo{volume}{25}}, \bibinfo{pages}{5677} (\bibinfo{year}{1992}).

\bibitem[{\citenamefont{Meystre and Wright}(1988)}]{Meystre1988a}
\bibinfo{author}{\bibfnamefont{P.}~\bibnamefont{Meystre}} \bibnamefont{and}
  \bibinfo{author}{\bibfnamefont{E.~M.} \bibnamefont{Wright}},
  \bibinfo{journal}{Physical Review A} \textbf{\bibinfo{volume}{37}},
  \bibinfo{pages}{2524} (\bibinfo{year}{1988}),
  \urlprefix\url{https://link.aps.org/doi/10.1103/PhysRevA.37.2524}.

\bibitem[{\citenamefont{Dalibard et~al.}(1992)\citenamefont{Dalibard, Castin,
  and M\o{}lmer}}]{Dalibard1992a}
\bibinfo{author}{\bibfnamefont{J.}~\bibnamefont{Dalibard}},
  \bibinfo{author}{\bibfnamefont{Y.}~\bibnamefont{Castin}}, \bibnamefont{and}
  \bibinfo{author}{\bibfnamefont{K.}~\bibnamefont{M\o{}lmer}},
  \bibinfo{journal}{Physical Review Letters} \textbf{\bibinfo{volume}{68}},
  \bibinfo{pages}{580} (\bibinfo{year}{1992}),
  \urlprefix\url{https://link.aps.org/doi/10.1103/PhysRevLett.68.580}.

\bibitem[{\citenamefont{Brun}(2002)}]{Brun2002a}
\bibinfo{author}{\bibfnamefont{T.~A.} \bibnamefont{Brun}},
  \bibinfo{journal}{American Journal of Physics} \textbf{\bibinfo{volume}{70}},
  \bibinfo{pages}{719–737} (\bibinfo{year}{2002}), ISSN
  \bibinfo{issn}{1943-2909},
  \urlprefix\url{http://dx.doi.org/10.1119/1.1475328}.

\bibitem[{\citenamefont{Gr{\"u}nbaum et~al.}(2013)\citenamefont{Gr{\"u}nbaum,
  Vel{\'a}zquez, Werner, and Werner}}]{Gruenbaum2013a}
\bibinfo{author}{\bibfnamefont{F.~A.} \bibnamefont{Gr{\"u}nbaum}},
  \bibinfo{author}{\bibfnamefont{L.}~\bibnamefont{Vel{\'a}zquez}},
  \bibinfo{author}{\bibfnamefont{A.~H.} \bibnamefont{Werner}},
  \bibnamefont{and} \bibinfo{author}{\bibfnamefont{R.~F.}
  \bibnamefont{Werner}}, \bibinfo{journal}{Communications in Mathematical
  Physics} \textbf{\bibinfo{volume}{320}}, \bibinfo{pages}{543}
  (\bibinfo{year}{2013}), ISSN \bibinfo{issn}{1432-0916},
  \urlprefix\url{http://dx.doi.org/10.1007/s00220-012-1645-2}.

\bibitem[{\citenamefont{Schulman}(1998)}]{Schulman1998a}
\bibinfo{author}{\bibfnamefont{L.~S.} \bibnamefont{Schulman}},
  \bibinfo{journal}{Physical Review A} \textbf{\bibinfo{volume}{57}},
  \bibinfo{pages}{1509} (\bibinfo{year}{1998}),
  \urlprefix\url{https://link.aps.org/doi/10.1103/PhysRevA.57.1509}.

\bibitem[{\citenamefont{Echanobe et~al.}(2008)\citenamefont{Echanobe, {del
  Campo}, and Muga}}]{Echanobe2008a}
\bibinfo{author}{\bibfnamefont{J.}~\bibnamefont{Echanobe}},
  \bibinfo{author}{\bibfnamefont{A.}~\bibnamefont{{del Campo}}},
  \bibnamefont{and} \bibinfo{author}{\bibfnamefont{J.~G.} \bibnamefont{Muga}},
  \bibinfo{journal}{Physical Review A} \textbf{\bibinfo{volume}{77}},
  \bibinfo{pages}{032112} (\bibinfo{year}{2008}).

\bibitem[{\citenamefont{Facchi and Pascazio}(2008)}]{Facchi2008a}
\bibinfo{author}{\bibfnamefont{P.}~\bibnamefont{Facchi}} \bibnamefont{and}
  \bibinfo{author}{\bibfnamefont{S.}~\bibnamefont{Pascazio}},
  \bibinfo{journal}{Journal of Physics A: Mathematical and Theoretical}
  \textbf{\bibinfo{volume}{41}}, \bibinfo{pages}{493001}
  (\bibinfo{year}{2008}).

\bibitem[{\citenamefont{Muga et~al.}(2008)\citenamefont{Muga, Echanobe, del
  Campo, and Lizuain}}]{Muga2008a}
\bibinfo{author}{\bibfnamefont{J.~G.} \bibnamefont{Muga}},
  \bibinfo{author}{\bibfnamefont{J.}~\bibnamefont{Echanobe}},
  \bibinfo{author}{\bibfnamefont{A.}~\bibnamefont{del Campo}},
  \bibnamefont{and} \bibinfo{author}{\bibfnamefont{I.}~\bibnamefont{Lizuain}},
  \bibinfo{journal}{Journal of Physics B: Atomic, Molecular and Optical
  Physics} \textbf{\bibinfo{volume}{41}}, \bibinfo{pages}{175501}
  (\bibinfo{year}{2008}),
  \urlprefix\url{https://doi.org/10.1088%2F0953-4075%2F41%2F17%2F175501}.

\bibitem[{\citenamefont{Sch{\"a}fer et~al.}(2014)\citenamefont{Sch{\"a}fer,
  Herrera, Cherukattil, Lovecchio, Cataliotti, Caruso, and
  Smerzi}}]{Schaefer2014a}
\bibinfo{author}{\bibfnamefont{F.}~\bibnamefont{Sch{\"a}fer}},
  \bibinfo{author}{\bibfnamefont{I.}~\bibnamefont{Herrera}},
  \bibinfo{author}{\bibfnamefont{S.}~\bibnamefont{Cherukattil}},
  \bibinfo{author}{\bibfnamefont{C.}~\bibnamefont{Lovecchio}},
  \bibinfo{author}{\bibfnamefont{F.~S.} \bibnamefont{Cataliotti}},
  \bibinfo{author}{\bibfnamefont{F.}~\bibnamefont{Caruso}}, \bibnamefont{and}
  \bibinfo{author}{\bibfnamefont{A.}~\bibnamefont{Smerzi}},
  \bibinfo{journal}{Nature communications} \textbf{\bibinfo{volume}{5}},
  \bibinfo{pages}{3194} (\bibinfo{year}{2014}).

\bibitem[{\citenamefont{Dhar et~al.}(2015{\natexlab{a}})\citenamefont{Dhar,
  Dasgupta, and Dhar}}]{Dhar2015a}
\bibinfo{author}{\bibfnamefont{S.}~\bibnamefont{Dhar}},
  \bibinfo{author}{\bibfnamefont{S.}~\bibnamefont{Dasgupta}}, \bibnamefont{and}
  \bibinfo{author}{\bibfnamefont{A.}~\bibnamefont{Dhar}},
  \bibinfo{journal}{Journal of Physics A: Mathematical and Theoretical}
  \textbf{\bibinfo{volume}{48}}, \bibinfo{pages}{115304}
  (\bibinfo{year}{2015}{\natexlab{a}}).

\bibitem[{\citenamefont{Dhar et~al.}(2015{\natexlab{b}})\citenamefont{Dhar,
  Dasgupta, Dhar, and Sen}}]{Dhar2015b}
\bibinfo{author}{\bibfnamefont{S.}~\bibnamefont{Dhar}},
  \bibinfo{author}{\bibfnamefont{S.}~\bibnamefont{Dasgupta}},
  \bibinfo{author}{\bibfnamefont{A.}~\bibnamefont{Dhar}}, \bibnamefont{and}
  \bibinfo{author}{\bibfnamefont{D.}~\bibnamefont{Sen}},
  \bibinfo{journal}{Physical Review A} \textbf{\bibinfo{volume}{91}},
  \bibinfo{pages}{062115} (\bibinfo{year}{2015}{\natexlab{b}}).

\bibitem[{\citenamefont{Elliott and Vedral}(2016)}]{Elliott2016a}
\bibinfo{author}{\bibfnamefont{T.~J.} \bibnamefont{Elliott}} \bibnamefont{and}
  \bibinfo{author}{\bibfnamefont{V.}~\bibnamefont{Vedral}},
  \bibinfo{journal}{Physical Review A} \textbf{\bibinfo{volume}{94}},
  \bibinfo{pages}{012118} (\bibinfo{year}{2016}),
  \urlprefix\url{https://link.aps.org/doi/10.1103/PhysRevA.94.012118}.

\bibitem[{\citenamefont{M{\"u}ller et~al.}(2017)\citenamefont{M{\"u}ller,
  Gherardini, and Caruso}}]{Mueller2017a}
\bibinfo{author}{\bibfnamefont{M.~M.} \bibnamefont{M{\"u}ller}},
  \bibinfo{author}{\bibfnamefont{S.}~\bibnamefont{Gherardini}},
  \bibnamefont{and} \bibinfo{author}{\bibfnamefont{F.}~\bibnamefont{Caruso}},
  \bibinfo{journal}{Annalen der Physik} \textbf{\bibinfo{volume}{529}},
  \bibinfo{pages}{1600206} (\bibinfo{year}{2017}),
  \eprint{https://onlinelibrary.wiley.com/doi/pdf/10.1002/andp.201600206}.

\bibitem[{\citenamefont{Lahiri and Dhar}(2019)}]{Lahiri2019a}
\bibinfo{author}{\bibfnamefont{S.}~\bibnamefont{Lahiri}} \bibnamefont{and}
  \bibinfo{author}{\bibfnamefont{A.}~\bibnamefont{Dhar}},
  \bibinfo{journal}{Physical Review A} \textbf{\bibinfo{volume}{99}},
  \bibinfo{pages}{012101} (\bibinfo{year}{2019}),
  \urlprefix\url{https://link.aps.org/doi/10.1103/PhysRevA.99.012101}.

\bibitem[{\citenamefont{Grynberg et~al.}(2010)\citenamefont{Grynberg, Aspect,
  and Fabre}}]{Grynberg2010a}
\bibinfo{author}{\bibfnamefont{G.}~\bibnamefont{Grynberg}},
  \bibinfo{author}{\bibfnamefont{A.}~\bibnamefont{Aspect}}, \bibnamefont{and}
  \bibinfo{author}{\bibfnamefont{C.}~\bibnamefont{Fabre}},
  \emph{\bibinfo{title}{Introduction to Quantum Optics}}
  (\bibinfo{publisher}{Cambridge University Press, New York},
  \bibinfo{year}{2010}), ISBN \bibinfo{isbn}{978-0-521-55112-0}.

\bibitem[{\citenamefont{Yin et~al.}(2019)\citenamefont{Yin, Ziegler, Thiel, and
  Barkai}}]{Yin2019a}
\bibinfo{author}{\bibfnamefont{R.}~\bibnamefont{Yin}},
  \bibinfo{author}{\bibfnamefont{K.}~\bibnamefont{Ziegler}},
  \bibinfo{author}{\bibfnamefont{F.}~\bibnamefont{Thiel}}, \bibnamefont{and}
  \bibinfo{author}{\bibfnamefont{E.}~\bibnamefont{Barkai}},
  \bibinfo{journal}{Physical Review Research} \textbf{\bibinfo{volume}{1}},
  \bibinfo{pages}{033086} (\bibinfo{year}{2019}),
  \urlprefix\url{https://link.aps.org/doi/10.1103/PhysRevResearch.1.033086}.

\bibitem[{\citenamefont{Liu et~al.}(2020)\citenamefont{Liu, Yin, Ziegler, and
  Barkai}}]{Liu2020a}
\bibinfo{author}{\bibfnamefont{Q.}~\bibnamefont{Liu}},
  \bibinfo{author}{\bibfnamefont{R.}~\bibnamefont{Yin}},
  \bibinfo{author}{\bibfnamefont{K.}~\bibnamefont{Ziegler}}, \bibnamefont{and}
  \bibinfo{author}{\bibfnamefont{E.}~\bibnamefont{Barkai}},
  \emph{\bibinfo{title}{Quantum walks: the first detected transition time}}
  (\bibinfo{year}{2020}), \eprint{arXiv:2001.00231}.

\bibitem[{\citenamefont{Thiel et~al.}(2019{\natexlab{a}})\citenamefont{Thiel,
  Mualem, Meidan, Barkai, and Kessler}}]{Stupid2019a}
\bibinfo{author}{\bibfnamefont{F.}~\bibnamefont{Thiel}},
  \bibinfo{author}{\bibfnamefont{I.}~\bibnamefont{Mualem}},
  \bibinfo{author}{\bibfnamefont{D.}~\bibnamefont{Meidan}},
  \bibinfo{author}{\bibfnamefont{E.}~\bibnamefont{Barkai}}, \bibnamefont{and}
  \bibinfo{author}{\bibfnamefont{D.~A.} \bibnamefont{Kessler}},
  \emph{\bibinfo{title}{Quantum total detection probability from repeated
  measurements i. the bright and dark states}}
  (\bibinfo{year}{2019}{\natexlab{a}}), \eprint{1906.08112}.

\bibitem[{\citenamefont{Thiel et~al.}(2019{\natexlab{b}})\citenamefont{Thiel,
  Mualem, Meidan, Barkai, and Kessler}}]{Thiel2020b}
\bibinfo{author}{\bibfnamefont{F.}~\bibnamefont{Thiel}},
  \bibinfo{author}{\bibfnamefont{I.}~\bibnamefont{Mualem}},
  \bibinfo{author}{\bibfnamefont{D.}~\bibnamefont{Meidan}},
  \bibinfo{author}{\bibfnamefont{E.}~\bibnamefont{Barkai}}, \bibnamefont{and}
  \bibinfo{author}{\bibfnamefont{D.~A.} \bibnamefont{Kessler}},
  \emph{\bibinfo{title}{Quantum total detection probability from repeated
  measurements}} (\bibinfo{year}{2019}{\natexlab{b}}), \bibinfo{note}{uploaded
  to ArXiv}, \eprint{1906.08112}.

\bibitem[{\citenamefont{Bourgain et~al.}(2014)\citenamefont{Bourgain,
  Gr{\"u}nbaum, Vel{\'a}zquez, and Wilkening}}]{Bourgain2014a}
\bibinfo{author}{\bibfnamefont{J.}~\bibnamefont{Bourgain}},
  \bibinfo{author}{\bibfnamefont{F.~A.} \bibnamefont{Gr{\"u}nbaum}},
  \bibinfo{author}{\bibfnamefont{L.}~\bibnamefont{Vel{\'a}zquez}},
  \bibnamefont{and}
  \bibinfo{author}{\bibfnamefont{J.}~\bibnamefont{Wilkening}},
  \bibinfo{journal}{Communications in Mathematical Physics}
  \textbf{\bibinfo{volume}{329}}, \bibinfo{pages}{1031} (\bibinfo{year}{2014}),
  ISSN \bibinfo{issn}{1432-0916},
  \urlprefix\url{https://doi.org/10.1007/s00220-014-1929-9}.

\end{thebibliography}
\end{document}